\documentclass[12pt,english]{article}
\usepackage[T1]{fontenc}
\usepackage[latin9]{inputenc}
\usepackage{geometry}
\usepackage{babel}
\usepackage{array}
\usepackage{float}
\usepackage{textcomp}
\usepackage{url}
\usepackage{multirow}
\usepackage{amsmath}
\usepackage{amsthm}
\usepackage{amssymb}
\usepackage{graphicx}
\usepackage[unicode=true]
 {hyperref}

\makeatletter

\newcommand{\noun}[1]{\textsc{#1}}
\providecommand{\tabularnewline}{\\}
\newcommand{\lyxdot}{.}

\theoremstyle{definition}
\newtheorem*{defn*}{\protect\definitionname}
\theoremstyle{remark}
\newtheorem*{rem*}{\protect\remarkname}

\usepackage{amsmath}
\usepackage{url}
\usepackage{listings}
\usepackage{url}

\makeatother

\providecommand{\definitionname}{Definition}
\providecommand{\remarkname}{Remark}

\begin{document}
\title{How to hunt wild constants}
\author{David R. Stoutemyer\thanks{Retired professor, Information and Computer Science Department, University
of Hawaii}}
\maketitle
\begin{abstract}
There are now several comprehensive web applications, stand-alone
computer programs and computer algebra functions that, given a floating
point number such as 6.518670730718491, can return concise nonfloat
constants such as $\,3\arctan2+\ln9+1\,$ that closely approximate
the float. This is analogous to unscrambling an egg. Such software
includes AskConstants, Inverse Symbolic Calculator, the Maple identify
function, MESearch, OEIS, Ordner, RIES, and WolframAlpha. Usefully
often such a result is the exact limit as the float is computed with
increasing precision. Therefore these program results are candidates
for proving an exact result that you could not derive or conjecture
without the program. Moreover, candidates that are \textsl{not} the
exact limit can be provable bounds, or convey qualitative insight,
or suggest series that they truncate, or provide sufficiently close
efficient approximations for subsequent computation. This article
describes some of these programs, how they work, and how best to use
each of them. Almost everyone who uses or should use mathematical
software can benefit from acquaintance with several such programs,
because these programs differ in the sets of constants that they can
return.
\end{abstract}

\section{Introduction}
\begin{flushright}
``\ldots\textsl{ you are in a state of }\,constant\,\textsl{ learning.}''\\
\textendash{} Bruce Lee
\par\end{flushright}

This article is about numerical \textsl{mathematical} constants that
can be computed approximately, rather than about dimensionless or
dimensioned physical constants. This article is a more detailed version
of the conference presentation \cite{StoutemyerConstantHunters}.

For real-world problems, we often cannot directly derive exact closed-form
results even with the help of computer algebra, but we more often
can compute approximate floating-point results \textendash{} hereinafter
called \textbf{floats}. For some such cases there is an exact closed-form
expression that the float approximates, and that expression is simple
enough so that we would like to know it, but we do not know how to
derive it or to guess it as a prerequisite to a proof. If we had one
or a few plausible concise nonfloat candidates that agree sufficiently
closely with the float, then we could concentrate our efforts on attempting
to prove that one of those candidates is the exact result.

This article describes text and software tools that provide such candidates.
Table 1 lists several such tools in order of presentation, grouped
by type. Figure 1 plots the initial publication dates of these tools
on a time line. All of the web-based and downloadable tools are free,
but some of the downloadable tools require Java, a C compiler, Maple
or \textsl{Mathematica}.\footnote{\textsl{Mathematica} is free and pre-installed on some Raspberry Pi
computers costing only a few dollars.}

\begin{table}[H]
\caption{Tables in books, web apps, standalone apps, CAS functions, and CAS
apps}

\noindent \begin{centering}
\smallskip{}
\par\end{centering}
\label{TableSoftwardAndPrinted}

\begin{tabular}{|c|c|c|c|c|}
\hline 
\textbf{Type} & \textbf{Name} & $\!$$\!$\textbf{Sec}.$\!$$\!$ & \textbf{Needs} & \textbf{http://www. or https://www.}\tabularnewline
\hline 
\hline 
\multirow{3}{*}{$\!\!\!\!\!$%
\begin{tabular}{c}
\textbf{Book}\tabularnewline
\textbf{table}\tabularnewline
\end{tabular}$\!\!\!\!\!$} & Robinson $\!$\&$\!$ Potter  & \ref{subsec:RobinsonPotter} & \multirow{3}{*}{{\footnotesize{}$\!\!\!\!\!$}%
\begin{tabular}{c}
{\footnotesize{}Library}\tabularnewline
{\footnotesize{}or}\tabularnewline
{\footnotesize{}buy}\tabularnewline
\end{tabular}{\footnotesize{}$\!\!\!\!\!$}} & {\footnotesize{}\href{http://escholarship.org/uc/item/2t95c0bp}{escholarship.org/uc/item/2t95c0bp}}\tabularnewline
\cline{2-3} \cline{3-3} \cline{5-5} 
 & $\!\!$Borwein $\!$\&$\!$ Borwein$\!\!$ & \ref{subsec:OtherTables} &  & {\footnotesize{}\href{http://springer.com/gp/book/9781461585121}{springer.com/gp/book/9781461585121}}\tabularnewline
\cline{2-3} \cline{3-3} \cline{5-5} 
 & Steven Finch & \ref{subsec:OtherTables} &  & {\footnotesize{}$\!$\href{http://sites.oxy.edu/lengyel/originals/0521818052ws.pdf}{sites.oxy.edu/lengyel/originals/0521818052ws.pdf}$\!$}\tabularnewline
\hline 
$\!\!\!\!\!$%
\begin{tabular}{c}
\textbf{.pdf}\tabularnewline
\textbf{table}\tabularnewline
\end{tabular}$\!\!\!\!\!$ & Shamo's Catalog & \ref{subsec:Shamos=002019s-Catalog} &  & {\footnotesize{}\cite{Shamos}}\tabularnewline
\cline{1-3} \cline{2-3} \cline{3-3} \cline{5-5} 
$\!\!\!\!\!$%
\begin{tabular}{c}
\textbf{Web}\tabularnewline
\textbf{search}\tabularnewline
\end{tabular}$\!\!\!\!\!$ & Google, etc. & \ref{sec:Web-browsers} & \multirow{2}{*}{{\footnotesize{}$\!\!\!\!\!$}%
\begin{tabular}{c}
{\footnotesize{}Smart-}\tabularnewline
{\footnotesize{}phone}\tabularnewline
\end{tabular}{\footnotesize{}$\!\!\!\!\!$}} & {\footnotesize{}\href{http://Google.com}{Google.com}, etc.}\tabularnewline
\cline{1-3} \cline{2-3} \cline{3-3} \cline{5-5} 
 & OEIS & \ref{subsec:The-On-line-Encyclopedia} &  & {\footnotesize{}\href{http://oeis.org}{oeis.org}}\tabularnewline
\cline{2-3} \cline{3-3} \cline{5-5} 
 & Ordner & \ref{subsec:Ordner} & {\footnotesize{}or} & {\footnotesize{}\href{http://fungrim.org/ordner}{fungrim.org/ordner}}\tabularnewline
\cline{2-3} \cline{3-3} \cline{5-5} 
$\!\!\!\!\!$\textbf{Web}$\!\!\!\!\!$ & WolframAlpha & \ref{subsec:WolframAlpha} & \multirow{2}{*}{{\footnotesize{}computer}} & {\footnotesize{}\href{http://wolframalpha.com}{wolframalpha.com}}\tabularnewline
\cline{2-3} \cline{3-3} \cline{5-5} 
$\!\!\!\!\!$\textbf{apps}$\!\!\!\!\!$ & ISC & \ref{subsec:Inverse-Symbolic-Calculator} &  & {\footnotesize{}$\!\!$\href{http://wayback.cecm.sfu.ca/projects/ISC/ISCmain.html}{wayback.cecm.sfu.ca/projects/ISC/ISCmain.html}$\!\!$}\tabularnewline
\cline{2-3} \cline{3-3} \cline{5-5} 
 & ISC+ & \ref{subsec:ISP+} &  & {\footnotesize{}\href{http://isc.carma.newcastle.edu.au/}{isc.carma.newcastle.edu.au/}}\tabularnewline
\hline 
\multirow{2}{*}{$\!\!\!\!\!$%
\begin{tabular}{c}
\textbf{Stand}\tabularnewline
\textbf{alone}\tabularnewline
\end{tabular}$\!\!\!\!\!$} & MESearch & \ref{subsec:MESearch} & {\footnotesize{}Java} & {\footnotesize{}\href{http://plouffe.fr/MESearch/}{http://plouffe.fr/MESearch/}}\tabularnewline
\cline{2-5} \cline{3-5} \cline{4-5} \cline{5-5} 
 & RIES & \ref{subsec:RIES} & {\footnotesize{}C compiler} & {\footnotesize{}\href{http://mrob.com/ries}{mrob.com/ries}}\tabularnewline
\hline 
\multirow{2}{*}{$\!\!\!\!\!$%
\begin{tabular}{c}
\textbf{In}\tabularnewline
\textbf{CAS}\tabularnewline
\end{tabular}$\!\!\!\!\!$} & identify & \ref{subsec:MapleIdentify-} & {\footnotesize{}Maple} & {\footnotesize{}\href{http://maplesoft.com/support/help/Maple/view.aspx}{maplesoft.com/support/help/Maple/view.aspx}}\tabularnewline
\cline{2-5} \cline{3-5} \cline{4-5} \cline{5-5} 
 & $\!$identify, $\!$findpoly$\!$ & \ref{subsec:The-MPMath,-SymPy} & {\footnotesize{}SymPy} & {\footnotesize{}\href{http://docs.sympy.org/0.7.1/modules/mpmath/}{docs.sympy.org/0.7.1/modules/mpmath/}}\tabularnewline
\hline 
\multirow{2}{*}{$\!\!\!\!\!$%
\begin{tabular}{c}
\textbf{CAS}\tabularnewline
\textbf{apps}\tabularnewline
\end{tabular}$\!\!\!\!\!$} & AskConstants & \ref{subsec:AskConstants} & {\footnotesize{}$\!$}\textsl{\footnotesize{}Mathematica}{\footnotesize{}$\!$} & {\footnotesize{}\href{http://AskConstants.org}{AskConstants.org}}\tabularnewline
\cline{2-5} \cline{3-5} \cline{4-5} \cline{5-5} 
 & Plouffe's Inverter & \ref{subsec:Inverter} & {\footnotesize{}Maple} & {\footnotesize{}\href{http://forthcoming:\%20\%20archive.org}{forthcoming:  archive.org}}\tabularnewline
\hline 
$\!\!\!\!\!$\textbf{\noun{PSLQ}}$\!\!\!\!\!$ & $\!$$\!$Custom IR model$\!$$\!$ & \ref{sec:Custom-Integer-Relation} & {\footnotesize{}$\!\!\!\!\!$}%
\begin{tabular}{c}
{\footnotesize{}Fortran}\tabularnewline
{\footnotesize{}C$_{++}$ or CAS}\tabularnewline
\end{tabular}{\footnotesize{}$\!\!\!\!\!$} & {\footnotesize{}\cite{Bailey,BaileyAndBroadhurst}}\tabularnewline
\hline 
\end{tabular}
\end{table}

\begin{figure}[H]
\caption{Tables, web apps, download apps and functions to propose exact constants}

\noindent \begin{centering}
\smallskip{}
\par\end{centering}
\label{FigTimeLine}

\includegraphics[clip]{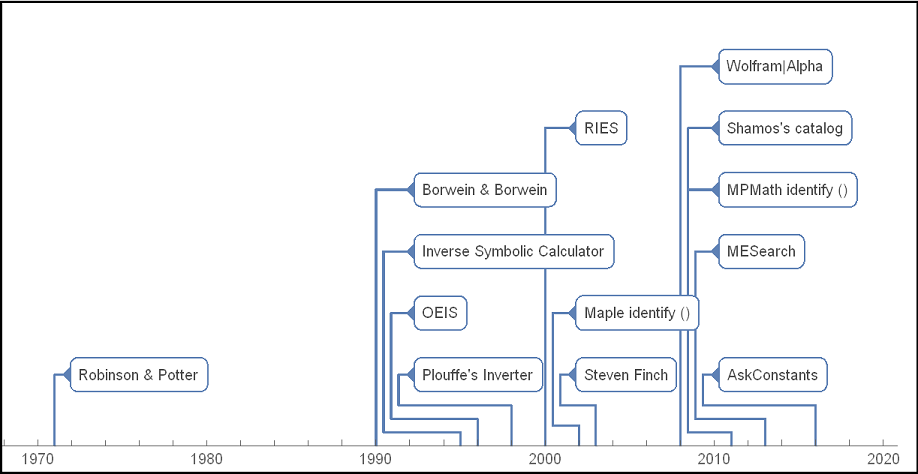}
\end{figure}

Float inputs for these tools often arise from computations such as
numerical integration, infinite series, iterative equation solving,
approximate optimization, etc. The purposes of this article are to
explain how best to use these tools, to explain how they work, and
to explain how to decide if proposed candidates are promising or else
probably impostors.

Section 8 discusses input magnitude limitations and Section 9 discusses
some common causes of impostor results, with conclusions in Section
10.

\section{Text tables\label{sec:Printed-tables}}

\subsection{\textsl{Mathematical Constants} table by Robinson \& Potter\label{subsec:RobinsonPotter}}

Here is an excerpt from a hand-typed table of about 3000 constants
by Robinson and Potter \cite{RobinsonAndPotter}:
\begin{center}
\smallskip{}
\par\end{center}

\begin{center}
\begin{tabular}{|c|c|}
\hline 
... & ...\tabularnewline
\hline 
4 .22755 35333 76265 40809 & $-\psi(1/4)=\gamma+3\ln2+\pi/2$\rule[-3pt]{0pt}{17pt}\tabularnewline
\hline 
0 .22755 09577 68849 99385 & 4/$\left(\pi^{2}e^{\gamma}\right)$\rule[-5pt]{0pt}{18pt}\tabularnewline
\hline 
0 .22756 34054 87472 14332 & root of $7xe^{x}=2$\rule[-4pt]{0pt}{17pt}\tabularnewline
\hline 
... & ...\tabularnewline
\hline 
\end{tabular}
\par\end{center}

\smallskip{}

Notice that:
\begin{itemize}
\item Inverse to more common tables of constants, the inputs in the left
column are floats and the expressions in the right column are corresponding
exact constants.
\item The digamma function of 1/4 was negated to make the float positive.
Most such tables do this \textbf{sign aliasing}, because it is easy
to discard the sign of the float, then negate the corresponding nonfloat
result. This doubles the potential coverage without increasing the
table size.
\item The inputs are sorted by the \textsl{fractional parts} of the absolute
values of the input floats rather than by those entire input floats.
For a given float $\tilde{x}$, you do a manual search for the fractional
part of $\left|\tilde{x}\right|$, then decide whether or not the
discrepancy with the fractional part of either bracketing entry or
of entries slightly further away is small enough to justify further
consideration. This \textbf{fractional-part aliasing} makes the table
applicable to many more examples and is easily inverted mentally to
construct the true candidate. For example, from the table float 4
.22755\,35333\,76265\,40809, we can also easily guess that a numerical
result 5.22755\,35332 might approach $1-\psi(1/4)$ as the precision
increases.
\item The last entry in the above table excerpt is an implicit result. Since
the popularization of the Lambert $W$ function \cite{LambertW},
this result can now be expressed explicitly as $W_{0}\,(2/7)$, but
the table contains approximate solutions to other equations that cannot
yet be expressed explicitly, together with approximate definite integrals,
infinite series, and infinite products having no known closed form.
\end{itemize}
To view a photocopy of the entire table, visit
\begin{center}
\url{https://escholarship.org/uc/item/2t95c0bp}
\par\end{center}
\begin{defn*}
\textbf{\textsl{Published constants}} are publicly accessible closed-form
and/or approximate float constants \textendash{} either printed or
on the web.
\end{defn*}
\vspace{-2pt}

\begin{defn*}
\textbf{\textsl{Named constants}} are constants having a widely-accepted
name, such as Catalan's constant or the \textsl{twin prime} constant.
Named constants include those with an OEIS name as described in subsection
\ref{subsec:The-On-line-Encyclopedia}.
\end{defn*}
\vspace{-2pt}

\begin{defn*}
\textbf{\textsl{Tabulated constants}} are formed by systematically
applying sets of functions to systematic sets of arguments, Printed
function tables for approximate computation usually have equally-spaced
arguments that are terminating decimal fractions such as 1.001, 1.002,
1.003, etc. In contrast, to return nonfloat results such as $\arctan(2/3)$,
the curated and computer-generated tables described in this article
often instead use the set of all reduced fraction arguments whose
numerator and denominator magnitudes do not exceed some given integer
\textendash{} perhaps also multiplied by common irrational constants
such as $\pi$, $\sqrt{2}$, etc. 
\end{defn*}
\vspace{-2pt}

\begin{rem*}
The table by Robinson and Potter contains all three of the these types.
\end{rem*}
\vspace{-2pt}

\begin{defn*}
\textsl{Wild constants} are computed float constants for which \textbf{you}
do not know an exact closed form for the limit as the precision of
the float approaches infinity.
\end{defn*}

\subsection{Other printed text tables\label{subsec:OtherTables}}
\begin{itemize}
\item Borwein and Borwein authored a similar printed table of about 100,000
constants \cite{Borweins}. The table contains mostly computer-generated
tabulated constants and some dimensionless physical constants.
\item The award-winning book by Steven Finch \cite{Finch} has not only
a table of about 10,000 well chosen non-tabulated constants, but it
and Volume II \cite{Finch2} contain short essays about those constants,
with references to the literature about them.
\end{itemize}

\section{Downloadable .pdf and web-searchable tables\label{sec:Shamos}}

\subsection{Shamos\textquoteright s Catalog of the Real Numbers\label{subsec:Shamos=002019s-Catalog}}

Michael Shamos posted a .pdf file of a table of about 10,000 constants
at \url{http://citeseerx.ist.psu.edu/viewdoc/download?doi=10.1.1.366.9997&rep=rep1&type=pdf}
Most of these constants are isolated pairs of a float value and a
closed form defined by a definite integral, infinite series, or infinite
product. Many floats list several corresponding formulas, such as
various integrals, series and products that have the same value, which
is helpful by suggesting other perhaps more efficient ways to compute
more digits.

A web search for ``\textbf{Mathematical constants}'' can locate
other tables that can be downloaded or searched directly on the web.

\section{Web browsers\label{sec:Web-browsers}}

General-purpose web browser search engines are also helpful for directly
finding candidate nonfloats for particular float inputs.

Here are some \textbf{input tips} for using the Google Search Engine
for this purpose\footnote{The Microsoft Edge and Apple Safari browsers currently allows a choice
of Google, Bing, Yahoo or DuckDuckGo as a search engine . }:
\begin{enumerate}
\item I infer that currently the search is based on your complete digit
string and complete digit substrings in a web site, rather than on
parsing your input float string and numeric web substrings into floats
then comparing those with a tolerance or on returning results for
which your input string matches a proper leading substring of a web
string or vice versa. This has several consequences:
\begin{enumerate}
\item You will probably obtain disjointly different sets of web sites for
different numbers of entered digits.
\item Entering too few digits is likely to overwhelm you with irrelevant
elementary mathematics tutorials, serial numbers, etc.
\item Entering too many digits is likely to return no result.
\item Most computed floats published since the internet are computed using
IEEE binary64 arithmetic, which is almost 16 digits, the last one
or more of which are often incorrect due to discretization, rounding
errors and catastrophic cancellation. Authors who realize this should
round or at least truncate their web numbers to at most the number
of digits they believe are almost certainly correct, but realization
of that is not a requirement for web publication. Thus if your number
was computed with IEEE binary 64, a reasonable strategy is to first
try all 16 digits, then 15, 14, etc. down to where you obtain promising
results or it seems likely than you will not obtain any. When doing
so, remember that an author might have rounded or truncated a number
on their web page, so you should try both if they differ. When rounding,
always do so from your complete original number because, for example,
rounding 1.546 to 1.55, then subsequently rounding that to 1.6 is
different than rounding 1.546 directly to 1.5$\,$.
\item But numbers of interest to mathematicians are often computed to more
than 16 digits, so if the next thing to try if your number was computed
to more than 16 digits is to try 17 digits, then 18, etc. Some authors
tend to prefer a multiple of 3, 4 or 5 for either the number of fractional
digits or the number of fractional digits \textendash{} particularly
if they group digits with spaces or commas for legibility, which might
affect the success of the web search.
\end{enumerate}
\item Decimal fraction notation seems more successful than scientific notation,
perhaps partly because of variations in spacing, characters, raised
exponents, and programming notations such as using E or e for ``$\times10$\textasciicircum ''.
However, if your float has magnitude that is quite large or small
compared to 1, then also try omitting the power of 10 from scientific
notation because the search engine might terminate a lexical scan
when it encounters any character other than a digit or decimal point
in a web page.
\item For negative numbers, try both the negative entry and its absolute
value.
\item For numbers having magnitude less than 1, try both with and without
a leading 0 digit before the decimal point.
\item If a web page containing a promising match includes a name for a constant
(such as ``Artin's constant''), then branch into a search using
that clue.
\end{enumerate}
Somewhat different tips might be appropriate for other search engines,
and Google's search algorithm might evolve to make some of the tips
here unnecessary. To learn, experiment with float approximations to
a few known mathematical constants that are not too obscure or too
well known. For example, try
\[
0.91596559\,.
\]

\section{Free web applications}

If you have internet access as you read this, then you might want
to try each of the web applications in this section as you read the
subsections. With good eyesight, most of them are usable even with
most smart phones.

To become acquainted with them, using simple then less simple nonfloat
constants from books and articles, try entering their floating-point
approximations of varying precisions. Also try random floats having
differing numbers of significant digits. As the number of significant
digits increases, random floats are decreasingly likely to be the
truncated or rounded value of any nonfloat expression simple enough
for any of these programs to propose in a reasonable amount of time.
An application is \textbf{generous} if it often suggests expressions
for such random-number entries, versus \textbf{parsimonious} if it
rarely does. It is alright for an application to be generous, but
it is important for you to realize it. Generosity can be helpful for
determining bounds or efficient sufficiently accurate approximations
or for suggesting the dominant terms of a longer exact result.
\noindent \begin{flushright}
``Perfection is the enemy of success''\\
\textendash{} Mark Cuban
\par\end{flushright}

Every float is exactly representable as an easily computed rational
number, but floats cannot exactly represent all rational numbers,
and floats are often approximations to irrational numbers. If your
float approaches a limit as the working precision of the calculation
approaches infinity and that limit is expressible as a nonfloat expression,
then that limit is what you want. \textbf{Perfect agreement with a
mere approximation prevents that desired result}.

\subsection{The On-line Encyclopedia of Integer Sequences\label{subsec:The-On-line-Encyclopedia}}

One of the main purposes of the OEIS site at \href{https://oeis.org/}{https://oeis.org/}
is to identify integer sequences such as proposing that the truncated
sequence 8, 16, 32, 64, . . . might be a subsequence of the sequence
whose nth element is $2^{n}$. Another goal of this database is to
provide accurate and comprehensive information about these sequences.

The successive digits of a decimal fraction can be regarded as an
integer sequence, and an under-advertised capability of OEIS is to
propose nonfloat constant expressions that closely approximate your
float. OEIS has sequences of up through about 100 digits or more for
about 10,000 constants regarded as important by the many contributors
who posted them. To attempt identifying nonfloat candidates for your
float, you can enter a comma-separated list of some significant digits
of the \textsl{absolute value} of its decimal fraction value, such
as 3, 1, 4, 1, 5, 9, 2, 6 . OEIS then attempts to match your entered
successive digits with \textsl{any} subsequence of its sequences.
It then lists information about each of its integer sequences that
contain your sequence as a subsequence. However, it is better to enter
your float digits as a positive decimal fraction with a decimal point,
such as 3.1415926. (Do not use scientific notation.)

Any matched table entries can contain hyperlinks to references and
contain computer-algebra program fragments to compute the constant
to any number of digits efficiently. There is also usually a Keyword
link labeled \textbf{cons} that leads to the precomputed value of
the constant up to a median of about 100 significant digits. The references
sometimes contain values with thousands of digits. 

This exact-match \textsl{decimal radix search} requires you to \textsl{truncate}
trailing digits of the absolute value of your float \textsl{rather
than round}. However, the hardware or software that computed your
float is approximate and almost certainly used rounding rather than
truncation. Therefore you should truncate enough of your trailing
digits to be reasonably confident that your submission matches exactly
the leading digits at OEIS.

If your computed float ends with a sequence of digits 9, then also
try what you get by adding 1 to the last digit. If your computed float
ends with a sequence of digits 0, then also try what you get by subtracting
1 from the last digit, because those minimal relative changes to your
approximate float can dramatically alter the digit sequence.

If you enter too few digits then you will obtain numerous matches,
most or all of which are irrelevant. However, with about $10^{5}$
numbers in the data base, entering 8 or more significant digits is
usually sufficient to achieve at most one match, and almost all of
the OEIS sequences for floats contain at least that many digits. Therefore
you can usually truncate to about 8 significant digits without much
chance of false matches or of entering more digits than are in a table
entry. However, some of the table entries are there because they are
surprisingly close to other constants of widespread interest, such
as $\pi$. Therefore you should always compare \textsl{all} of your
digits with corresponding ones stored at the cons link in the table.

Moreover, if you do obtain more than one match, then it is wise to
do this comparison for all of them or to enter more correctly truncated
digits if possible to reduce the match count to a comfortably inspectable
number. 

Since OEIS ignores the position of the decimal point, entering 31.41592
also returns $\pi$ rather than $10\pi$. It is your responsibility
to notice this and to multiply the proposed exact constant by the
appropriate power of 10. This base-ten\textbf{ significand aliasing}
increases the number of floats that can be matched.

Since OEIS can match any \textsl{subsequence} of their digit sequences,
you can also exploit fractional part aliasing by entering only the
fractional part of the absolute value of your float. For example,
if your truncated wild float is 9.141592654, then entering it produces
no result, whereas entering 0.141592654 produces $\pi$ and gives
its float value, enabling you to recognize that your wild float is
probably $6+\pi$. Thus it is worth trying all of your float magnitude,
then only the integer part of that magnitude if different.

Beware that fractional part aliasing suffers catastrophic cancellation
when the absolute value of the fractional part is small compared to
the absolute value of the float, whereas significand aliasing suffers
no loss of precision. Therefore it is best if 8 or more significant
digits remain after omitting the integer part and truncating dubious
trailing digits.

Other than you exploiting these types of aliasing, OEIS makes no effort
to identify more general transformations of its constants, such as
2/3 of a named constant or 2/3 plus a named constant.

All OEIS integer sequences and float digit sequences are assigned
an OEIS name consisting of the letter ``A'' followed by six digits.
Although such names are difficult to remember and do not acknowledge
the discoverers, they are unique and serve as a \emph{de facto} standard
\textendash{} particularly when there is no other widely recognized
name. In contrast, traditional names can have various spellings, subsets,
and orderings of all the discoverers, as well as values differing
by a factor of 2, etc, on account of one author using a radius where
another uses a diameter, for example.

A complimentary way to use OEIS for named constants is to enter a
name, such as \textsl{grazing goat}, or an OEIS name such as A133731.
If successful, you will find the same information as for entering
the truncated float value 1.158728473 .

Brief sorted descriptions of all the constants together with links
to their pages are at
\begin{center}
\url{https://oeis.org/wiki/Index_to_constants#Description}
\par\end{center}

Some of these constants are tabulated, but most are isolated published
constants, drawn from many areas by many contributors.

Even if another tool has already returned a satisfying result, OEIS
and Steven Finch's books are good places to seek more information
about a constant.

\subsection{Ordner: index of real numbers\label{subsec:Ordner}}

Fredrik Johansson \cite{JohanssonsOrdner} posted an online table
of over 3000 sign-aliased real numbers, many of which I have not found
elsewhere. To avoid a tedious amount of scrolling, there is a clickable
index of intervals that transfer you to the subtable bracketing the
absolute value of your float. For each subtable float there are one
or more links to alternative equivalent expressions having that 30-digit
value to within one unit in the last place.

As a bonus, a link to the Fungrim home page reveals a table of contents
with links into a collection of useful function identities, some of
which I have not seen elsewhere.

\subsection{WolframAlpha\label{subsec:WolframAlpha}}

If you enter a float into WolframAlpha at \url{https://www.wolframalpha.com/},
the results usually include up to three guesses for a nonfloat constant
that the float approximates.

A way to test the tools described in this article is to approximate
a nonfloat constant to a generous number of significant digits, then
see how many of them are necessary to produce an equivalent expression,
if any. Accordingly, I approximated
\begin{equation}
1-\dfrac{3\sqrt{3}}{1+\sqrt{3}+\pi}\label{eq:nonfloatForAlpha}
\end{equation}
to 18 digits, then tried entering that and successively fewer digits
into WolframAlpha. Table \ref{TableAlphaResults} lists the results
for floats \textsl{rounded} to 10 and to 11 significant digits, together
with entries having 11 correct digits followed by one and by two incorrect
digits. Underlined digits in the entries differ from the exact expression
by at most 0.5 units in the last place.

\begin{table}[H]
\caption{WolframAlpha test for identifying $1-\dfrac{3\sqrt{3}}{1+\sqrt{3}+\pi}$
from a float approximation\rule[-14pt]{0pt}{30pt}}

\noindent \begin{centering}
\label{TableAlphaResults}
\par\end{centering}
\centering{}%
\begin{tabular}{|c|c|c|c|}
\hline 
\textbf{Entry} & $\!$$\!$\textbf{digits}$\!$$\!$ & $\!$$\!$\textbf{Result (Agrees with input to 0.5 in last bold digit)}$\!$$\!$ & $\!$$\!$\textbf{$\boldsymbol{\equiv}$}$\!$$\!$\tabularnewline
\hline 
\hline 
\multirow{3}{*}{$\!$$\!$$\!$$\!$0.$\underline{1153442566}$} & \multirow{3}{*}{10} & $\dfrac{11-7\pi+2\pi^{2}}{\pi(21+\pi)}\approx\mathbf{0.115344256}49366$\rule[-7pt]{0pt}{28pt} & \tabularnewline
\cline{3-4} \cline{4-4} 
 &  & %
\begin{tabular}{r}
$\dfrac{1}{\mathrm{root\:of}\;x^{4}-75x^{2}+10x-99\;\mathrm{near}\;x=8.6697}$\rule[-5pt]{0pt}{26pt}\tabularnewline
$\approx\mathbf{0.115344256}56239$\rule[-6pt]{0pt}{16pt}\tabularnewline
\end{tabular} & \tabularnewline
\cline{3-4} \cline{4-4} 
 &  & %
\begin{tabular}{r}
$\mathrm{root\:of}\;99x^{4}-10x^{3}+75x^{2}-1\;\mathrm{near}\;x=0.115344$\rule[-4pt]{0pt}{17pt}\tabularnewline
$\approx\mathbf{0.11534425}\mathbf{6}56239$\tabularnewline
\end{tabular} & \tabularnewline
\hline 
\multirow{3}{*}{$\!$$\!$$\!$0.$\underline{11534425658}$} & \multirow{3}{*}{11} & $\dfrac{1-2\sqrt{3}+\pi}{1+\sqrt{3}+\pi}\approx\mathbf{0.11534425658}4835$\rule[-10pt]{0pt}{30pt} & $\!$$\!$$\boldsymbol{\boldsymbol{\checkmark}}$$\!$$\!$\tabularnewline
\cline{3-4} \cline{4-4} 
 &  & $\dfrac{-2e\,e!+124-35e+5e^{2}}{136e}\approx\mathbf{0.1153442565}79706$\rule[-11pt]{0pt}{32pt} & \tabularnewline
\cline{3-4} \cline{4-4} 
 &  & %
\begin{tabular}{r}
$\mathrm{root\:of}\;92x^{4}+49x^{3}-27x^{2}+37x-4\;\mathrm{near}\;x=0.115344$\rule[-4pt]{0pt}{17pt}\tabularnewline
$\approx\mathbf{0.11534425658}21563$\rule[-4pt]{0pt}{13pt}\tabularnewline
\end{tabular} & \tabularnewline
\hline 
 &  & $\dfrac{1-2\sqrt{3}+\pi}{1+\sqrt{3}+\pi}\approx\mathbf{0.115344256581}4835$\rule[-10pt]{0pt}{30pt} & $\!$$\!$$\boldsymbol{\boldsymbol{\checkmark}}$$\!$$\!$\tabularnewline
\cline{3-4} \cline{4-4} 
$\!$0.$\underline{11534425658}$0 & 12 & %
\begin{tabular}{c}
$\dfrac{2(15\sigma_{S}-1)}{5(6\sigma_{S}+73)}\approx\mathbf{0.11534425658}18196$\rule[-7pt]{0pt}{28pt}\tabularnewline
where $\sigma_{S}$ is Somos's quadratic recurrence constant\rule[-5pt]{0pt}{13pt}\tabularnewline
\end{tabular} & \tabularnewline
\cline{3-4} \cline{4-4} 
 &  & %
\begin{tabular}{r}
$\!$$\!$$\pi\left(\mathrm{\;root\:of}\;100x^{4}-152x^{3}-x^{2}-27x+1\;\mathrm{near}\;x=0.0367\right)$$\!$$\!$\rule[-4pt]{0pt}{17pt}\tabularnewline
$\approx\mathbf{0.11534425658}28298\rule[-4pt]{0pt}{13pt}$\tabularnewline
\end{tabular} & \tabularnewline
\hline 
\multirow{3}{*}{$\!$0.$\underline{11534425658}$00$\!$$\!$} & \multirow{3}{*}{13} & $\dfrac{45419\pi}{1237062}\approx\mathbf{0.1153442565800}221712$\rule[-10pt]{0pt}{29pt} & \tabularnewline
\cline{3-4} \cline{4-4} 
 &  & $\dfrac{3e\,e!+295-86e-9e^{2}}{94e}\approx\mathbf{0.153442565}79947$\rule[-11pt]{0pt}{32pt} & \tabularnewline
\cline{3-4} \cline{4-4} 
 &  & %
\begin{tabular}{r}
$\dfrac{1}{45}\left(4C+30-11\pi-40\pi\log(2)+27\pi\log(3)\right)$\rule[-6pt]{0pt}{25pt}\tabularnewline
$\!$$\!$$\mathbf{0.1153442565800}78601$, where $C$ is Catalan's
constant$\!$$\!$\tabularnewline
\end{tabular} & \tabularnewline
\hline 
\end{tabular}
\end{table}

Notice that:
\begin{enumerate}
\item Alpha did not propose an equivalent for 10 correct digits.
\item Alpha proposed an equivalent expression for 11 or more correct digits
provided they were followed by no more than 1 incorrect digit.
\item Thus Alpha almost always finds expressions that differ from your input
by at most a few units in the last place that you entered, making
those differences not a good criteria for favoring one alternative
over the others.
\item The returned candidates are usually nonequivalent, so Alpha is generous.
\end{enumerate}
So with typically three nonequivalent expressions all having about
the same differences from your input, how are we to know which if
any are the limit we seek?

By the principle of parsimony, you should prefer the least complex,
but the three alternatives for 11 correctly-rounded digits don't visually
differ much in complexity. Here is a way to gain some evidence: If
you can compute your wild float to more digits, perhaps using Alpha,
then do so and round the result enough to be reasonably confident
that the last digit or at least the one before it is correctly rounded,
then enter that float. Impostors for the lower-precision input are
less likely to appear for the higher-precision float. For example,
the correctly rounded 18-digit value for expression (\ref{eq:nonfloatForAlpha})
is 0.115344256581483524, and entering that returns the three alternatives

\[
\begin{alignedat}{1}\dfrac{1-2\sqrt{3}+\pi}{1+\sqrt{3}+\pi} & \approx\mathbf{0.1153442565814835244}14074,\\
\dfrac{26534199\pi}{722703038} & \approx\mathbf{0.11534425658148352}51297,\\
\dfrac{-678+46e+97e^{2}}{2\left(-10+167e+36e^{2}\right)} & \approx\mathbf{0.11534425658148352}36159\,.
\end{alignedat}
\]
Notice that the first nonfloat is the same as for entering 11 or 12
correctly rounded digits in Table \ref{TableAlphaResults}. The other
two nonfloats for 11 and 12 digits have been replaced with other expressions
and were therefore almost certainly impostors. It is possible that
the true limit would not appear until more than 18-digit precision,
making the first entry for 18 digits an impostor too that would disappear
at higher precision. However, it is typical of true limits that once
they appear at a certain precision, they will appear for all higher
precisions. Therefore it is advisable to enter increasing numbers
of probably-correctly rounded digits until one of the alternatives
convincingly persists or you become convinced that Alpha does not
model your float. If two or more candidates persist, then perhaps
they are equivalent.

If you cannot compute more digits for your original float, then try
rounding it to fewer digits to see which alternatives persist in that
direction.

When Alpha returns more than one result and only one has dramatically
less complexity than the others, there is some justification for preferring
it. However, I prefer to \textbf{always tray t least two idfferent
precisions to asses transcience versus persistence}.

Alpha mostly uses the PSLQ \textbf{integer relation} algorithm\footnote{The LLL algorithm can also be used for this purpose.}
\cite{Ferguson and Bailey,FergusonBaileyAndArno}: Given a basis vector
$\mathbf{c}=[c_{1},c_{2},$$\ldots]$ of nonzero symbolic and/or float
constants, the algorithm returns either an indication of failure or
a minimal 2-norm vector of integers $[n_{1},n_{2},\ldots]$ not all
0 such that the inner product $n_{1}c_{1}+n_{2}c_{2}+\cdots\simeq0.0$
and $\gcd(n_{1},n_{2},\ldots)=1\,.$
\begin{enumerate}
\item One way to exploit this algorithm is to make $c_{1}$ be the given
float $\tilde{x}$ and choose commonly occurring nonfloat constants
such as 1, $\pi$, $\sqrt{2}$, and $\ln2$ as the other constants.
If the algorithm returns a vector $[n_{1},n_{2},\ldots]$ , then a
close-fitting candidate is the \textsl{rational linear combination}
\[
\tilde{x}\simeq-\,\dfrac{n_{2}}{n_{1}}c_{2}-\dfrac{n_{3}}{n_{1}}c_{3}-\cdots.
\]
The last entry in Table \ref{TableAlphaResults} is of this type.
\item Another way to exploit this algorithm is if
\[
\left(n_{1}c_{1}+n_{2}c_{2}+\cdots\right)\tilde{x}+\left(n_{k}c_{k}+n_{k+1}c_{k+1}+\cdots\right)\simeq0.0,
\]
then
\[
\tilde{x}\simeq-\,\dfrac{n_{1}c_{1}+n_{2}c_{2}+\cdots}{n_{k}c_{k}+n_{k+1}c_{k+1}+\cdots}.
\]
This \textsl{linear fractional} category of model was used for the
correct identifications and some of the incorrect ones in Table \ref{TableAlphaResults}.
\item Another way to exploit this algorithm is if
\begin{equation}
n_{1}+n_{2}\tilde{x}+\cdots+n_{m+1}\tilde{x}^{m}\simeq0.0\label{eq:MinimalPolynomial}
\end{equation}
then $\tilde{x}$ is approximately one of the zeros of the polynomial
$n_{1}+n_{2}x+\cdots+n_{m+1}x^{m}$. This can be tried for $m=1,2,\ldots$
up to some maximum degree that increases with the number of significant
digits. If PSLQ returns a vector of integers, then an approximate
polynomial zero algorithm can then be used to determine which zero
is closest to the given float. Several alternatives in Table \ref{TableAlphaResults}
are such an \textsl{algebraic number} candidates.
\item The minimal polynomial integer relation can be generalized to include
predetermined exact irrational constants with one or more powers of
$\tilde{x}$, such as
\[
n_{1}\pi+n_{2}\tilde{x}+n_{2}\tilde{x}^{2}\simeq0.0\,.
\]
For degree 2 models such as this, the quadratic formula could be used
to express a successful PSLQ candidate result explicitly in terms
of $\pi$, rather than implicitly using Root {[}\ldots{]}.
\item For a power product model of the form $\tilde{x}\simeq b_{1}^{r_{1}}b_{2}^{r_{2}}\cdots$
with given positive nonfloat base constants $b_{k}$ and unknown rational
exponents $r_{k}$, if $n_{1}\ln\tilde{x}+n_{2}\ln b_{1}+n_{3}\ln b_{2}\cdots\simeq0.0$,
then
\[
\tilde{x}\simeq\exp\left(-\dfrac{n_{2}\ln b_{1}+n_{3}\ln b_{2}+\cdots}{n_{1}}\right)=b_{1}^{-n_{2}/n_{1}}b_{2}^{-n_{3}/n_{1}}\cdots.
\]
The irrational factors of an irrational exponent can be included in
the bases to make this category fit a model such as $\tilde{x}\simeq2^{r_{1}}3^{r_{2}\sqrt{5}}e^{r_{3}\pi}=2^{r_{1}}\left(3^{\sqrt{5}}\right)^{r_{2}}\left(e^{\pi}\right)^{r_{3}}$
.
\item For a model of the form $f(\ldots)$, where $f$ is invertible with
respect to at least one of its arguments and that argument model is
one of the above types or recursively of this functional type:
\begin{itemize}
\item Using predetermined constants for any other arguments, apply an inverse
of $f$ to the given float $\tilde{x}$ giving $\tilde{y}$.
\item If the argument model is successfully fit to $\tilde{y}$, giving
a nonfloat candidate constant $y$, then return $f(y)$.
\end{itemize}
\end{enumerate}
The computing time grows rapidly with the number of elements $m$
in the basis vector and the number of words in the float significands.
Moreover, the PSLQ algorithm can require the input float to have $mn$
significant digits of precision to return a correct $m$ component
vector having integer components of up through $n$ digits even if
many of these integer coefficients have much smaller magnitudes, including
0$\,$. Therefore, although specialized efforts to identify particular
floats have used thousands of digits and hundreds of vector components,
the general-purpose programs described here use mostly basis vectors
having $\lesssim8$ elements.

\paragraph{More advise for using Alpha to identify floats:}
\begin{enumerate}
\item If you enter 18 or more significant digits, including any trailing
0 digits, then Alpha uses arbitrary precision rather than 16-digit
machine floats. Arbitrary-precision floats are slower but more than
18 digits are necessary for PSLQ to determine many published constants.
\item Alternatively, you can append a suffix of the form $`n$ to the significand
of your input to indicate that your best guess for the Precision is
the value of $n$, which can be a decimal fraction. For example, if
your input for the example in Table \ref{TableAlphaResults} was computed
by iterative equation solving using 16-digit IEEE binary64 hardware
and you guess that most probably about 14 digits are correctly rounded,
then you could start by entering 0.1153442565814834`14 . This \textsl{strongly
recommended} alternative has the advantage that you can enter \textsl{all}
of the computed digits, which which are more likely to be correct
than the value rounded or truncated to 14 digits, but without having
Alpha assume that all but perhaps the last one are correct. In the
other direction, if your computed 16-digit machine float is 0.1615000000000000
and you are rather confident that all are correct, then you can enter
it more concisely and reliably as 0.1615`16 . Different from the apostrophe,
the accent grave character ` is near the upper left corner on many
keyboards and might look different on your keyboard or screen.
\item Software does not always display all of the digits in its significands,
and internet browsers paste into web applications such as Alpha only
the \textsl{characters} that you highlight when copying. For example,
\textsl{Mathematica} displays by default only 6 of the approximately
16 significant digits in the widespread IEEE binary64 machine floats,
and displays none of the 8 or more stored \textsl{guard} digits in
arbitrary-precision floats. Often some of those hardware or guard
digits are correct and therefore helpful to try entering. Therefore,
learn how to display all of the digits in your software's floats.\footnote{One way to do this in \textsl{Mathematica} is to copy all of the displayed
digits then paste into a \textsl{notebook}, then copy all of that
fully-displayed InputForm float into Alpha. Another way is to copy
the result of $\mathrm{InputForm}\,[\mathit{float}]$ into Alpha.}
\item Unless you are an exceptionally fast accurate digit typist, highlight
then \textbf{copy the entire float including its hidden digits rather
than type it}.
\item Then attach to the significand a suffix of the form `$n$, where $n$
is your best guess for the number of significant digits that are correctly
rounded. Keep in mind that cancellation and rounding errors often
lose one or more digits from the working precision \textendash{} and
that discretization errors such as from quadrature often lose more.
Interval arithmetic can provide an upper bound on cancellation and
rounding errors, but not necessarily on discretization errors. Significance
arithmetic such as \textsl{Mathematica's} arbitrary-precision float
arithmetic provides cancellation and rounding error estimates that
are occasionally underestimates but more frequently overestimates. 
\item PSLQ overfits often contain many digits in their rational numbers,
such as the first alternative in Table \ref{TableAlphaResults} for
13 digits. Information theory and my experience suggests that results
are dubious if they contain a total of more instances of operators,
functions and digits than about 6 less than your number of estimated
correct significant digits in your float input. Also, several terms
or factors containing only simple rational numbers is less suspicious
than one term containing a very complicated rational number.
\end{enumerate}

\subsection{Inverse Symbolic Calculator\label{subsec:Inverse-Symbolic-Calculator}}

Inverse Symbolic Calculator (ISC) implemented by Simon Plouffe working
with Jonathon and Peter Borwein and others is currently hosted at
\begin{center}
\url{http://wayback.cecm.sfu.ca/projects/ISC/ISCmain.html} .
\par\end{center}

\begin{flushleft}
Plouffe \cite{PlouffeISCHistory} describes its origins. 
\par\end{flushleft}

ISC provides four different techniques to apply to the float that
you enter:

\subsubsection{Simple Lookup and Browser}

This technique searches a precomputed table of sorted \textbf{truncated}
16-digit base-ten \textbf{significands} of the \textbf{absolute values}
of the floats, each paired with a character-string description of
the constant.

That table has about 30 million pairs that are a mixture of published
and tabulated constants. The sorted significands and the corresponding
nonfloats are subdivided into about 9000 files whose names include
some leading digits of all the significands in that file, with each
file containing all of the significands having those leading digits.
The search begins with a binary search on the file names in RAM, followed
by a search in the appropriate file to perfectly match or bracket
the significand of the float that you entered, followed by retrieval
of close neighbors of those entries. Here are some \textbf{input tips}:
\begin{itemize}
\item \begin{flushleft}
ISC Simple Lookup defines a \textbf{match} as \textbf{all} of the
entered digits of the significand being \textbf{identical} to that
leading portion of a tabulated significand, but ignoring any entered
digits beyond 16.
\par\end{flushleft}
\item The last one or more digits of your computed float are often not the
correctly truncated ones. To maximize your chance of success, try
to compute as many digits as is practical to be reasonably confident
that the first 16 digits are correctly truncated, then enter at least
those 16 digits. Your float was probably computed with software that
rounds rather than truncates, so at least truncate the last computed
digit.
\item If you cannot compute more than 16 digits, then truncate to as many
as your best guess for how many are correctly truncated digits of
the limit you seek.
\item If the least significant truncated digit is a 9, then also try truncating
what you get by adding 1 to that 9 digit. If the least significant
truncated digit is instead a 0, then also try truncating what you
get by subtracting 1 from that 0 digit. Those minimal relative changes
to your entered float can dramatically alter the trailing digit sequence.
\item You can enter as few as 5 significant digits, but with about 30 million
entries having significands 0.1000000000000000 through 0.9999999999999999,
the mean distance between table significands is 
\[
\dfrac{0.9999999999999999-0.1000000000000000}{30\times10^{6}}\simeq3.3\times10^{-8}.
\]
Thus you can expect an unwelcome number of matches if you enter fewer
than seven or eight significant digits.
\item You are offered the opportunity to browse nearby table entries even
if there are no matches. Doing so might reveal one or more candidate
nonfloats that agree closely and are plausible considering your knowledge
of the problem domain. Browse even if there are matches, because your
input might not be as correctly truncated as you assumed.
\item The table contains many instances of nonfloats that all have the same
16-digit significand value. The corresponding nonfloats are often
but not always equivalent to each other within a sign and a factor
that is a power of 10. Approximating these nonfloats to higher precision
can support such aliased equivalence or reveal the lack of such equivalence.
\item Moreover, if you can compute your wild float to more than 16 digits
and you have software that can evaluate a table nonfloat to more than
16 digits, then doing so permits you to determine how many more digits
of that nonfloat agree with your higher-precision wild float. Beware
though that the table nonfloats are character strings some of which
are Maple syntax or \textsl{Mathematica} syntax or neither, so you
might have to type a syntactically appropriate translation into your
software rather than merely copy and paste.\footnote{Or, try typing ``N {[}\textsl{nonfloat}, \textsl{desiredPrecision}{]}''
into WolframAlpha, where \textsl{nonfloat} is the string copied and
pasted from ISC. WolframAlpha is somewhat tolerant of varying mathematical
notation.} 
\item If you do not obtain at least one promising result, then also try
entering fewer and more truncated digits up through 16, covering what
you guess spans most of the possible actual correctly truncated digits
of your wild float.
\end{itemize}
As a test example, the correctly truncated 16-digit value of
\begin{equation}
-10\sqrt{\dfrac{2}{37}\left(8\sqrt{3}-9\right)}=-5.123557917376186\,.\label{eq:SimpleLookupEg}
\end{equation}
Table \ref{TableISCResults} summarizes the results of entering various
further truncated and rounded leading digits to expose the consequences
of rounding rather than truncating \textendash{} and of appending
incorrectly truncated digits beyond the sixteenth.
\begin{table}[H]
\noindent \begin{centering}
\caption{ISC Simple Lookup test results for approximated $\left|-10\sqrt{\dfrac{2}{7}\left(8\sqrt{3}-9\right)}\right|$}
\rule[-8pt]{0pt}{18pt}\label{TableISCResults}
\par\end{centering}
\begin{tabular}{|l|c|c|c|c|}
\hline 
Input & $\!\!$digits$\!\!$ & $\!\!\!$Treatment$\!\!\!$ & Result & ?\tabularnewline
\hline 
\hline 
$5.1235$ & \multirow{2}{*}{5} & $\!\!\!$\textbf{truncated}$\!\!\!$ & $\!\!\!$\textbf{38 matches} include $12^{1/4}/\sqrt{8+9^{3/4}}$$\!\!\!$ & $\!\!\!$$\begin{array}{c}
\mathrm{Too}\\
\mathrm{many}
\end{array}$$\!\!\!$\tabularnewline
\cline{1-1} \cline{3-5} \cline{4-5} \cline{5-5} 
$5.1236$ &  & rounded & $\!$$\!$%
\begin{tabular}{c}
\textbf{360 matches}.\tabularnewline
They \textbf{don't} include $12^{1/4}/\sqrt{8+9^{3/4}}$\tabularnewline
\end{tabular}$\!$$\!$ & $\!\!\!$$\begin{array}{c}
\mathrm{Too}\\
\mathrm{many}
\end{array}$$\!\!\!$\tabularnewline
\hline 
$5.123557$ & \multirow{2}{*}{7} & $\!\!\!$\textbf{truncated}$\!\!\!$ & Matches only $12^{1/4}/\sqrt{8+9^{3/4}}$ & $\boldsymbol{\checkmark}$+\tabularnewline
\cline{1-1} \cline{3-5} \cline{4-5} \cline{5-5} 
$5.123558$ &  & rounded & %
\begin{tabular}{c}
6 matches that \textbf{do not}\tabularnewline
include nearby $12^{1/4}/\sqrt{8+9^{3/4}}$\tabularnewline
\end{tabular}$\!\!\!$ & $\!\!\!$$\begin{array}{c}
\mathrm{All}\\
\mathrm{bad}
\end{array}$$\!\!\!$\tabularnewline
\hline 
$5.1235579173$ & \multirow{2}{*}{11} & $\!\!\!$\textbf{truncated}$\!\!\!$ & Matches only $12^{1/4}/\sqrt{8+9^{3/4}}$ & $\boldsymbol{\checkmark}$+\tabularnewline
\cline{1-1} \cline{3-5} \cline{4-5} \cline{5-5} 
$5.1235579174$ &  & rounded & $\!\!\!$%
\begin{tabular}{c}
No match, but browser lists adjacent\tabularnewline
$12^{1/4}/\sqrt{8+9^{3/4}}$, matching 10 digits\tabularnewline
\end{tabular}$\!\!\!$ & $\boldsymbol{\checkmark}$\tabularnewline
\hline 
$5.12355791730$ & 12 & $\!\!\!$%
\begin{tabular}{c}
incorrect\tabularnewline
last digit\tabularnewline
\end{tabular}$\!\!\!$ & $\!\!\!$%
\begin{tabular}{c}
No match, but browser lists adjacent\tabularnewline
$12^{1/4}/\sqrt{8+9^{3/4}}$, matching 10 digits\tabularnewline
\end{tabular}$\!\!\!$ & $\boldsymbol{\checkmark}$\tabularnewline
\hline 
$5.12355791737618$ & \multirow{2}{*}{15} & $\!\!\!$\textbf{truncated}$\!\!\!$ & Matches only $12^{1/4}/\sqrt{8+9^{3/4}}$ & $\boldsymbol{\checkmark}$+\tabularnewline
\cline{1-1} \cline{3-5} \cline{4-5} \cline{5-5} 
$5.12355791737619$ &  & rounded & $\!\!\!$%
\begin{tabular}{c}
No match, but browser lists adjacent\tabularnewline
$12^{1/4}/\sqrt{8+9^{3/4}}$, matching 14 digits\tabularnewline
\end{tabular}$\!\!\!$ & $\boldsymbol{\checkmark}$\tabularnewline
\hline 
$5.1235579173761861$ & \multirow{2}{*}{17} & $\!\!\!$%
\begin{tabular}{c}
\textbf{incorrect}\tabularnewline
\textbf{last digit}\tabularnewline
\end{tabular}$\!\!\!$ & Matches only $12^{1/4}/\sqrt{8+9^{3/4}}$ & $\boldsymbol{\checkmark}$+\tabularnewline
\cline{1-1} \cline{3-5} \cline{4-5} \cline{5-5} 
$5.1235579173761869$ &  & $\!\!\!$%
\begin{tabular}{c}
\textbf{incorrect}\tabularnewline
\textbf{last digit}\tabularnewline
\end{tabular}$\!\!\!$ & Matches only $12^{1/4}/\sqrt{8+9^{3/4}}$ & $\boldsymbol{\checkmark}$+\tabularnewline
\hline 
\end{tabular}

\end{table}

For each match, ISC lists the truncated 16-digit base-10 significand
without a decimal point, equated to a corresponding nonfloat, such
as
\begin{equation}
512355791737618=12\wedge(1/4)/(8+9\wedge(3/4))\wedge(1/2)\,.\label{eq:SimpleLookupResult}
\end{equation}

To process this match:
\begin{enumerate}
\item The ratio of your input 5.123557917376186 to the significand 0.5123557917376186
is 10, so multiply the candidate nonfloat by 10 to de-alias the significand.
\item Then negate that, because our float of interest in equation (\ref{eq:SimpleLookupEg})
is negative and we necessarily entered its absolute value, giving
our final result
\begin{equation}
-10\,\dfrac{12^{1/4}}{\sqrt{8+9^{3/4}}}.\label{eq:FinalSimpleLookupResult}
\end{equation}
\item For verification of our de-aliasing and of our interpretation of the
table nonfloat, we should approximate expression (\ref{eq:FinalSimpleLookupResult})
to preferably more than 16 digits, then truncate to trusted digits,
giving -5.123557917376186, which agrees with the right side of equation
(\ref{eq:SimpleLookupEg}) to 16 digits.
\item It is not difficult to transform the de-aliased nonfloat candidate
(\ref{eq:FinalSimpleLookupResult} into our test expression on the
left side of (\ref{eq:SimpleLookupEg}). But what if our input was
a wild constant that happens to agree with an impostor to 16 digits?\textbf{}\\
\textbf{}\\
\textbf{For candidates resulting from tables, if you can compute your
float input and a candidate nonfloat to higher precision than used
in the table, then it is wise to assess exact and near-exact matches
to higher precision than used in the table to help expose impostors
or gain confidence in the correctness of the nonfloat.}
\item However, beware that the table nonfloats are character strings some
of which are Maple or \textsl{Mathematica} syntax or neither. Therefore
you might have to type a syntactically appropriate translation into
your software rather than merely copy and paste.\footnote{Or, try exploiting Alpha's heuristic expression parser by typing ``N
{[}\textsl{nonfloat}, \textsl{desiredPrecision}{]}'', where \textsl{nonfloat}
is the string copied and pasted from ISC.} 
\end{enumerate}
Beware also that the table was assembled over many years from many
computed and printed sources that used varying syntactic conventions.
For example:
\begin{itemize}
\item Gam, Gamma, GAM and GAMMA all denote the Gamma function, whereas gamma
denotes the Euler-Mascheroni constant.
\item sqrt and sr both denote the principal square root.
\item A polynomial in $x$ or$\,$ root(\textsl{a\_polynomial\_in\_x}) $\,$denotes
the zero of that polynomial nearest the given float.
\item Non-real expressions often implicitly denote the real parts of those
expressions.
\item As with our regrettably ambiguous traditional mathematics notation,
sometimes 16\textasciicircum 3/4 means 16\textasciicircum (3/4)
rather than (16\textasciicircum 3)/4, $\sin e+3$ means $\sin(e+3)$
rather than $\sin(e)+3$, and $7/3\pi$ means $7/(3\pi)$ rather than
$(7/3)\pi$, etc. 
\end{itemize}
\begin{quote}
\textbf{Step 3 above enables you to verify or refute your interpretation
of the nonfloat syntax, then to try another interpretation if necessary}\textsl{.}
\end{quote}
Some nonfloats in the table contain either a common name constant
such as Bernstein or a red OEIS web link labeled with an OEIS constant
having a name of the form ``A'' with a six-digit suffix. Either
way, you can search for their definitions at \url{https://oeis.org/},
as described in subsection \ref{subsec:The-On-line-Encyclopedia}.

\subsubsection{Smart Lookup\label{subsec:Smart-Lookup}}

This technique uses the same lookup table about 100 times, but precedes
each lookup with a transformation of your float. For example, smart
lookup might add 3/7 or multiply by 3/7 or apply $\ln(\ldots)$. For
a given number of input digits, the average number of matches is thus
about 100 times as many as for simple lookup However, some transformations
lose precision. Therefore smart lookup requires you to enter at least
10 digits, and the transformation of your input is done using 26-digit
significands. Thus it is beneficial to enter floats having up through
26 correctly rounded digits even though the forward table floats have
only 16-digit significands. 

For example, computed to 26 significant digits,
\begin{equation}
-\dfrac{1}{4}-10\,\dfrac{12^{1/4}}{\sqrt{8+9^{3/4}}}\simeq-5.3735579173761866715453232\,.\label{eq:SmartLookupEg}
\end{equation}

Applying simple lookup to the absolute value of this float truncated
to 16 digits returns no match, and the ISC browser lists no result
that agrees to more than 5 digits. However, applying \textsl{smart
lookup} to the absolute value of the right side of equation (\ref{eq:SmartLookupEg}),
computed with arithmetic that rounds, returns the table 
\noindent \begin{center}
\begin{tabular}{|c|c|c|c|}
\hline 
\textsl{Function} & \textsl{Result} & \textsl{Precision} & \textsl{Matches}\tabularnewline
\hline 
\hline 
$\mathrm{K}-1/4$ & 5.1235579173761866715453232 & 16 & 1\tabularnewline
\hline 
\end{tabular}
\par\end{center}

The \textsl{Function} entry, $\mathrm{K}-1/4$ is a transformation
that was applied to your float, with K representing both the nonfloat
that you seek and its approximate value
\[
5.3735579173761866715453232\,.
\]
The \textsl{Result} entry is the result of applying the transformation
to your float as computed with 26-digit significands. The \textsl{Precision}
entry is the number of leading digits in the table that matched those
in your transformed float. The \textsl{Matches} entry is the number
of matches for that transformation.

If you click on the \textsl{Function} entry $\mathrm{K}-1/4$, then
ISC shows the result of doing \textsl{simple lookup} with a truncated
significand of the transformed \textsl{Result} float, which gives

\begin{equation}
\dfrac{12^{1/4}}{\sqrt{8+9^{3/4}}},\label{formulaForK}
\end{equation}
together with an opportunity to browse around it.

If you want to further investigate one of these nonfloats that have
agreement acceptable to you, then, for example, you should
\begin{enumerate}
\item Numerically approximate the nonfloat, preferably to at least 16 digits,
giving
\[
\mathrm{K=}12^{1/4}/\sqrt{8+9^{3/4}}\simeq0.512355791737618667\,.
\]
\item From the ratio of that float to the float in the Result entry, determine
a power of 10 by which to multiply the nonfloat:
\[
\dfrac{5.1235579173761866715453232}{0.512355791737618667}=10.0000000000000000\,.
\]
\item Multiply the table nonfloat (\ref{formulaForK}) by that power of
10 giving the magnitude de-aliased nonfloat
\[
y=10\,\dfrac{12^{1/4}}{\sqrt{8+9^{3/4}}}
\]
\item Solve $y=\mathrm{K}-1/4$ for K either manually or by computer algebra
:), giving
\[
K=y+1/4=10\dfrac{12^{1/4}}{\sqrt{8+9^{3/4}}}.+\dfrac{1}{4}
\]
\item Negate this result to de-alias the sign, giving the nonfloat candidate
for our float of interest
\[
-\left(\dfrac{1}{4}+10\dfrac{12^{1/4}}{\sqrt{8+9^{3/4}}}\right).
\]
\end{enumerate}
To summarize: 
\begin{enumerate}
\item We made the input float positive by negating it.
\item Smart lookup found a transformation that lead to a matching significand
in the Simple lookup table.
\item We multiplied the corresponding nonfloat $y_{1}$ in the Simple lookup
table by an appropriate power of 10 to make the value of the resulting
nonfloat $y_{2}$ match the \textsl{Result} value in the Smart lookup
table.
\item We then applied the inverse of the \textsl{Function} of K in the Smart
lookup table to $y_{2}$ giving $y_{3}$.
\item We then negated $y_{3}$ giving the candidate for our original float.
\end{enumerate}
More generally there might be several rows in the returned table for
several different transformations that resulted in matches. (The number
of rows tends to decrease as you increase the number of entered digits
above the minimum required 10 digits.)

Click on each of the Function entries to see the corresponding nonfloats
and optionally browse their neighbors.

For this example, entering fewer than 16 digits produced no smart
lookup result.

It is worth starting with Simple Lookup first because it is faster,
it does not require an inverse transformation, and it permits browsing
the neighborhood of your input even if there are no matches.

\subsubsection{ISC Integer Relation Algorithms}

ISC provides an Integer Relation alternative that you should try too
even if simple or smart lookup gives a promising result. Integer Relations
requires inputs of 16 through 32 significant digits. Integer Relation
checks for algebraic numbers through fifth degree and for rational
linear combinations of the basis vectors
\begin{center}
\begin{tabular}{c}
$\left[e,\pi,\gamma,\mathrm{Ei}\,(1),W\,(1),1\right],$\rule[-8pt]{0pt}{18pt}\tabularnewline
$\left[\sqrt{3}\pi,\ln3,\ln2,\gamma,\sqrt{2}\pi\right],$\rule[-8pt]{0pt}{18pt}\tabularnewline
$\left[\pi^{2},\mathrm{Catalan},\pi\ln2,\sqrt{2}\pi^{2},(\ln2)^{2}\right],$\rule[-8pt]{0pt}{18pt}\tabularnewline
$\left[\pi^{3},\zeta\,(3),\pi^{2}\ln2,(\ln2)^{3},\sqrt{3}\pi^{3},\sqrt{2}\pi^{3}\right].$\rule[-8pt]{0pt}{18pt}\tabularnewline
\end{tabular}
\par\end{center}

These basis vectors might seem rather few and specialized. However,
a noticeable portion of wild constants in the printed tables of Section
\ref{sec:Printed-tables} are representable with a subset of at least
one of these vectors or with algebraic numbers of degree $\leq5$.

It is wise to compute as many digits as you can up through 32. There
is no need to truncate, and it is OK for a few of the last digits
to be incorrectly rounded.

\subsubsection{Generalized Expansions}

You must enter at least 16 significant digits for this option, and
it is most appropriate to \textsl{round} your approximate float to
where you are reasonably confident that the last digit is correctly
rounded.

This option computes a truncated continued fraction from your input
float, and computes similar truncated infinite representations such
as a truncated infinite product and a truncated Egyptian fraction.
This can be helpful because these alternatives might suggest patterns
that might be useful for computing the float to arbitrarily many digits
more efficiently than the method you used. Moreover, the GFUN package
\cite{GFUN} is then applied to these truncated representations to
try guessing a generating function for the infinite representation.
(It is for you to prove that it \textsl{is} such a generating function.)

Also, you can copy the comma-separated sequence of integers for one
of these truncated representations, then paste it at \href{https://oeis.org/}{https://oeis.org/}
to attempt learning more about it, possibly including a closed form.
The {[}Hints{]} button on the OEIS home page invokes advice that includes:
\begin{quote}
``Enter about 6 terms, starting with the second term. Leave off the
first term or two, because people may disagree about where the sequence
begins. Don't enter too many terms, because you may have more terms
than are in the OEIS data base.''
\end{quote}
Generalized Expansions and two related tools GCF.TXT by Dougherty-Bliss
and Zeilberger \cite{DoughertyBlissZeilberger} and the Ramanujan
Machine by a team of nine authors \cite{RamanujanMachine} are intriguing
but adjacent to the topic of this article, so they are not described
further here.

\subsection{Inverse Symbolic Calculator Plus\label{subsec:ISP+}}

Inverse Symbolic Calculator Plus at \href{https://isc.carma.newcastle.edu.au/}{https://isc.carma.newcastle.edu.au/}
was a different interface to essentially the same smart lookup and
integer-relations models as ISC, ISC+ has had the message ``down
for maintenance indefinitely'' from late 2018 through at least January
2022. However, as of this writing there is a link there titled ``The
original ISC'' that links to the ISC site already discussed in subsection
\ref{subsec:Inverse-Symbolic-Calculator}.

\section{Freely downloadable standalone applications\label{sec:Freely-downloadable-standalone}}

Most of the implemented integer relation models and precomputed table-lookup
entries of the software described in this article are based on one
implementer's perception of what patterns of nonfloat constant expressions
are most likely to fit the combined needs of the intended users. There
is a consequent efficiency to the extent that the implementer's choices
are appropriate for you. However, such an implementation is almost
certain to miss some very concise results fitting unimplemented patterns.

The MESearch and RIES software address this issue by trying all possible
expressions composed of user-selected rational numbers, symbolic constants,
operators and functions \textendash{} up to a certain total complexity,
time limit or memory limit. By default both programs use \textsl{bidirectional
search}: The \textbf{forward table} is like the ISC table, and the
\textbf{backward table} is like the ISC Smart Lookup transformations.
However, rather than using \textsl{precomputed} forward and transformation
tables, for each given float both applications build new tables from
your selected components in a breadth-first way, interleaving the
searching with this table building. The forward table begins with
a selected set of simple rational numbers and named constants together
with a set of arithmetic operators and functions. More complicated
forward expressions are formed by applying all chosen functions and
operators to existing entries in an order such that the predicted
expression complexity grows approximately monotonically. The backward
table starts with the given float, and applies the selected functions
in an order such that the estimated resulting complexity also grows
approximately monotonically. For functions and operators of more than
one operand, all but one of the operands can be taken from the current
forward table.

Growth alternates between the forward and backward tables so that
at any one time they have roughly the same number of entries.

Even though the initial set of rational numbers is typically quite
simple, arithmetic combinations of rational subexpressions can lead
to larger numerator and/or denominator magnitudes.

Both programs use only hardware floating point but make the best of
that limitation. Moreover, as with the 16-digit Inverse Symbolic Calculator
table, if you can approximate nonfloat candidates and your wild float
to greater than hardware precision, then you can assess the agreement
of candidates to greater precision to help distinguish impostors from
the limit you seek.

Both programs use a few simplification rules such as employing 0 and
1 identities, doing rational arithmetic, and canceling composition
of a function with its inverse. However, neither program uses a computer
algebra system, so sometimes the results can be usefully simplified
manually or by computer algebra.

Computation ceases at your preset expression complexity limit, memory
limit, time limit, or the limit of your patience when you finally
interrupt a search. Both programs initially use RAM for the tables,
but can resort to a much slower secondary-storage mode when allocated
RAM is exhausted.

The fact that the tables are exhaustively built and searched in approximate
order of increasing expression complexity has a benefit that simple
candidates composed of your selected components are almost always
quickly found. However, the open-ended exponential growth of the table
sizes and computing time with expression complexity makes it impractical
to generate expressions having complexity that is easily achievable
by sparse precomputed tables such as OEIS and ISC or by integer relation
models with sufficiently large-precision inputs. With a medium sized
set of selected components, exhaustive search can require overnight
to generate an expression containing about 12 or 13 instances of modest
rational numbers, symbolic constants, arithmetic operators, and function
invocations. 

But the peace of mind knowing that the search was exhaustive for expressions
composed of your selected rational numbers, symbolic constants, operators
and functions is worth the wait. (This is a good way for you to extract
value from your computer even while you sleep!)

In contrast, the other programs discussed in this article have a fixed
number of table entries and/or integer relation models. The latter
can accommodate expressions having any number of predetermined symbolic
constants in one of several categories of forms , with arbitrarily
large integers determined by PSLQ. But being non-exhaustive, those
tools can and do miss some very simple results that MESearch and/or
RIES can determine.

Applying some of the transformations to the given float might produce
unusable underflow overflow, or non-real results; and compositions
of different backward with forward pairs might lead to identical or
equivalent results. However, if the backward table has $m$ entries
and the forward table has $n$ entries, then a useful upper bound
on the number of potentially recognizable expressions is $mn$. If
both kinds of table entries averaged the same memory space per entry,
then for a given amount of memory, $mn$ is maximized by having $m=n$.
For example, only $10^{6}$ entries in the forward table and $10^{6}$
entries in the backward table gives an upper bound of recognizing
$10^{12}$ different-valued expressions.

MESearch and RIES both offer the option of using only a forward table,
but they are almost always faster and less memory consumptive bidirectionally.

\subsection{MESearch\label{subsec:MESearch}}

MESearch is implemented in Java and was freely downloadable from
\begin{center}
\url{http://www.xuru.org/mesearch/MESearch.asp}\,
\par\end{center}

That much changed site no longer lists MESearch, but thanks to Simon
Plouffe, a copy is freely downloadable from
\noindent \begin{center}
\url{http://plouffe.fr/MESearch/}
\par\end{center}

The Java Runtime Environment is often already installed on most computers,
and it is freely downloadable from
\noindent \begin{center}
\url{https://www.java.com/en/download/}
\par\end{center}

MESearch uses only invertible functions or operators for the top level
function or operator of its backward table, and automatically applies
those inverses to the nonfloats in the forward table whose float values
closely match a transformed float input. 

\begin{figure}[H]
\caption{Input panel for an MESearch example}
\label{FigMESearchInput}
\noindent \centering{}\includegraphics[clip]{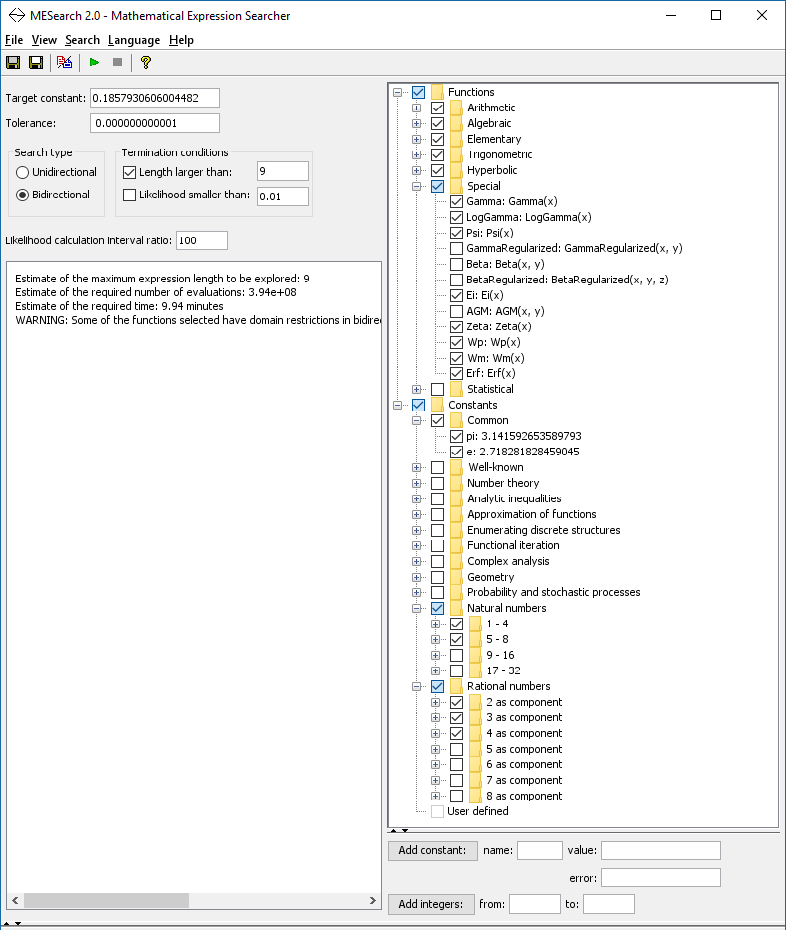}
\end{figure}

Figure \ref{FigMESearchInput} shows the MESearch input panel for
the test float input
\begin{equation}
\dfrac{1}{\sqrt{5+2\sqrt{6}}+\sqrt{5}}\approx0.1857930606004482.\label{eq:MESearchInputFloat}
\end{equation}

MESearch measures expression complexity by \textbf{Length}, defined
as the number of occurrences of rational numbers, named constants,
functions and operators. Counting the implicit multiply, the Length
of the expression on the left side of equation (\ref{eq:MESearchInputFloat})
is 12.

MESearch offers a choice of operators, functions, named constants,
natural numbers and rational numbers from which to build expressions,
and Figure \ref{FigMESearchInput} shows my most common choices for
these. Choosing only a few components permits you to search to longer
expressions before exhausting memory or your patience, but it is important
to include components that you expect have a nonnegligible probability
of occurring in your problem domain. Functions and operators with
more than one argument are particularly costly, and that is why I
often initially limit those to ``+'', ``$\times$'', and ``\textasciicircum ''.
Memory and computing time also grow quickly with the cardinality of
the initial set of named constants, natural numbers, and rational
numbers. Thus I often initially select only $\pi$ and $e$ together
with very simple natural numbers and fractions.

MESearch only displays candidates whose float value are within $\pm$Tolerance
entered under the Target constant near the upper left corner of Figure
2. The default MESearch Tolerance is 1/2 unit in the last entered
digit of your Target constant, which is difficult to achieve if your
Target constant is all 16 digits of the hardware floats or nearly
so. Consequently I often initially choose about 1000 to 10,000 units
in the last entered place for Target floats having 16 significant
digits, down to about 10 units in the last place for Target floats
having 7 or fewer significant digits. Generous tolerances change the
result likelihood estimates and are more likely to generate impostors,
but unlike overly frugal tolerances, generous tolerances do not tend
to lose achievable true limits. I would rather obtain a true limit
together with a \textsl{few} impostors, then decide which are most
plausible and recompute those with smaller tolerances for more accurate
likelihood estimates.

MESearch uses your entered \textbf{absolute tolerance} rather than
the number of entered digits. Therefore you should enter all of your
computed digits if there are less than 16 or rounded to 16 digits
otherwise, even if any number of the resulting trailing digits are
dubious. Therefore I entered all 16 digits of 0.1857930606004482,
for which my entered absolute tolerance was $10^{-12}$.

Here is a summary of the most important information in the resulting
output:
\begin{flushleft}
\begin{tabular}{|c|c|c|c|c|c|}
\hline 
$\!$$\!${\footnotesize{}Length}$\!$$\!$ & Expression & $\!$$\!$$\!${\footnotesize{}Difference}$\!$$\!$$\!$ & $\!$$\!$$\!${\footnotesize{}Likely}$_{1}$$\!$$\!$$\!$ & $\!$$\!$$\!${\footnotesize{}Likely}$_{2}$$\!$$\!$$\!$ & $\!$$\!$$\!${\footnotesize{}Likely}$_{3}$$\!$$\!$$\!$\tabularnewline
\hline 
\hline 
8 & ((e\textasciicircum arccosh$\,$(5))\textasciicircum (1/2) + 5\textasciicircum (1/2))\textasciicircum -1 & {\footnotesize{}0.0} & {\footnotesize{}0.5103} & {\footnotesize{}0.9135} & {\footnotesize{}0.9187}\tabularnewline
\hline 
9 & $\!$arctanh$\,$(cos$\,$(tan$\,$(sin$\,$(4/3$\,$/$\,$sinh$\,$(cos$\,$(4))))))$\!$ & {\footnotesize{}6.769e-13} & {\footnotesize{}3.3e-11} & {\footnotesize{}0.8515} & \tabularnewline
\hline 
9 & $\!$$\!$arctan($\psi$(2 - tan(Ei(logGamma((3/4)\textasciicircum 2)))))$\!$$\!$ & {\footnotesize{}6.769e-13} & {\footnotesize{}3.3e-11} & {\footnotesize{}0.8515 } & \tabularnewline
\hline 
\end{tabular}
\par\end{flushleft}

Watching the output panel evolve provides some entertainment while
you wait and helps you decide whether and when to terminate the search
manually.

The first expression that met the tolerance was at complexity Length
8. The three Likelihood estimates are based on the fact that the expressions
are all possible expressions composed of the selected components,
computed in approximately monotonic non-decreasing order of complexity.
The reason for the two vacant Likely$_{3}$ entries is that Likely$_{3}$
can be computed only upon \textbf{completion} of searching its Length
level, and I manually halted the search during Length 9.

Clicking a column heading sorts the rows by the values in that column,
so I usually click the middle Likely$_{2}$, which the implementer
Salsamendi states is most often most correct.

If this input float was a wild float, then having the 0.0 Difference
between the entered float and the value of the Length 8 Expression,
together with its two likelihoods of greater than 90\% and one of
greater than 50\% would lead me to strongly suspect that the corresponding
expression is a correct limit. In contrast, the very low values for
half the likelihoods of the two Level 9 candidates and their significantly
greater Differences would lead me to strongly suspect that they are
impostors.

The result subexpression $\,$e\textasciicircum arccosh$\,$(5)$\,$
having Length 4 is quite different from the corresponding test subexpression
$\,\sqrt{5+2\sqrt{6}}\,$ in equation (\ref{eq:MESearchInputFloat}).
However, they are provably equivalent, and MESearch almost certainly
would have generated the more recognizable equivalent
\[
\left(\sqrt{5+2\sqrt{6}}+\sqrt{5}\right)^{-1}
\]
at Level 11.

Although most people prefer algebraic numbers to be expressed using
radicals or trigonometric functions if possible and practical, other
representations often have lower complexity by most quickly-computed
measures.\footnote{Many computer algebra systems have transformation functions that can
help you seek alternative representations of result candidate expressions.}

Salsamendi \cite{SalsamendiArticle} describes the data structures,
algorithm, and time complexity.

\subsection{RIES (Rilybot Inverse Equation Solver)\label{subsec:RIES}}

RIES is implemented in C, and the source code is freely downloadable
from \url{https://mrob.com/ries} . Your computer might already have
a C or C++ compiler that you are unaware of, and there are several
good free ones for Mac OS X, the Unix family, and Windows.

RIES optionally exploits 19-digit IEEE binary80 arithmetic if your
CPU and C compiler support it. The resulting 19 digits is a valuable
increment over the typical 16-digit hardware. IEEE binary80 is supported
by Intel x86 processors, but not by current or past AMD or ARM processors.
The Intel C++ compiler and the free GNU C compiler support this arithmetic,
and Microsoft Visual C can be made to do so at the inconvenience of
inserting \_controlfp\_s function calls at appropriate places in the
source code.

Here is an example by Bill Gosper (private communication):

\textsl{Mathematica} 12.1 cannot determine a closed form for
\[
\sum_{n=0}^{\infty}\dfrac{(-1)^{n}\cos\left(\sqrt{\frac{1}{2}+\left(\frac{1}{2}+n\right)^{2}}\,\pi\right)}{\left(\frac{1}{2}+n\right)\left(\frac{1}{2}-\frac{1}{\sqrt{2}}+n\right)\left(\frac{1}{2}+\frac{1}{\sqrt{2}}+n\right)}
\]

Suppose that you compute a 20-digit approximate value 7.0895773641597344051,
and you invoke RIES as the command line
\begin{center}
\textsf{ries $\:$-l7 $\:$7.0895773641597344051 $\:$-s}
\par\end{center}

\noindent \begin{flushleft}
where option \textsf{-l7} specifies the maximum search level and option
\textsf{-s} permits implicit equational results that are not of the
explicit form \textsf{x = }\textsl{constantExpression}\textsf{. }This
feature is valuable because such equations are often solvable manually
or by computer algebra, and you at least have a possibly useful different
characterization of your wild constant that might provide insight
and enable you to approximate it to larger precision more efficiently.
\par\end{flushleft}

Here is the corresponding slightly edited RIES output: 

$\vphantom{}$

\textsf{Your target value: T = 7.08957736415973441 $\:$mrob.com/ries}

$\vphantom{}$

\textsf{x = pi + 4 \ for\  x = T + 0.0520153 $\:$\{49\}}

\textsf{x = 4 sqrt(pi) \ for\  x = T + 0.000238039 $\:$\{58\}}

\textsf{x = 1/(8 + pi) + 7 \ for\  x = T + 0.000176411 $\:$\{81\}}

\textsf{x = (sqrt(pi) + pi)/ln(2) \ for\  x = T - 0.000106841 $\:$\{87\}}

\textsf{x = e\textasciicircum (e\textasciicircum sqrt(phi))/5 \ for\ 
x = T - 8.60481e-05 $\:$\{90\}}

\textsf{x = ln(ln(6\textasciicircum 8))\textasciicircum 2 $\,$for$\,$
x = T + 7.23949e-05 $\:$\{93\}}

\textsf{x = (phi\textasciicircum 2)\textasciicircum sqrt(1 + pi)
$\,$for$\,$ x = T - 4.72474e-05 $\:$\{85\}}

\textsf{x = (e\textasciicircum (pi - 2))\textasciicircum 2 - e $\,$for$\,$
x = T + 1.25979e-05 $\:$\{90\}}

\textsf{x = (1/pi + sqrt(pi))\textasciicircum 2 + e $\,$for$\,$
x = T - 2.53137e-06 $\:$\{92\}}

\textsf{x = e (sinpi(1/ln(9)) + phi) $\,$for$\,$ x = T + 2.0796e-06
$\:$\{109\}}

\textsf{x = 3/pi + 8 - pi\textquotedbl /x $\,$for$\,$ x = T + 5.08817e-10
$\:$\{113\}}

\textsf{sinpi(sqrt(x)/e) = (1/(8/phi - 1))\textasciicircum 2 $\,$for$\,$
x = T - 3.68034e-10 $\:$\{131\}}

\textsf{x = 3((pi\textquotedbl /4 - 1/7) + phi) - 2 $\,$for$\,$
x = T - 3.51782e-11 $\:$\{141\}}

\textsf{x = 2 e\textquotedbl /(1 + e) - 1/(1 - 3\textquotedbl /2)
$\,$for$\,$ x = T + 1.65079e-11 $\:$\{147\}}

\textsf{x = ((log\_9(1/2 + 8))/4)/e + 7 $\,$for$\,$ x = T - 1.3897e-11
\{150\}}

\textsf{x = -(pi/cospi(1/sqrt(2))) - pi cospi(1/sqrt(2)) $\,$for$\,$
x = T - 5.28657e-16 $\:$\{148\}}

\textsf{x = pi (-1/cospi(1/sqrt(2)) - cospi(1/sqrt(2))) $\,$for$\,$
x = T + 3.05745e-16 $\:$\{154\}}

\textsf{x = (1/tanpi(1/sqrt(8))\textasciicircum 4 + 1)(x/2 + pi)
$\,$for$\,$ x = T - 1.30104e-18 $\:$\{164\}}

\textsf{$\vphantom{}$}

\textsf{(Stopping now because best match is within 3.07e-18 of target
value.)}

\textsf{$\vphantom{}$}

\textsf{log\_A(B) = logarithm to base A of B = ln(B) / ln(A),}

\textsf{cospi(X) = cos(pi {*} x)}

\textsf{sinpi(X) = sin(pi {*} x)}

\textsf{tanpi(X) = tan(pi {*} x)}

\textsf{phi = the golden ratio = (1+sqrt(5))/2}

\textsf{A\textquotedbl /B = Ath root of B}

\textsf{$\vphantom{}$}

\textsf{}%
\begin{tabular}{rrrrl}
 & \textsf{-{}-LHS-{}-} & \textsf{-{}-RHS-{}-} & \textsf{-Total-} & \tabularnewline
\textsf{max complexity:} & \textsf{86} & \textsf{81} & \textsf{167} & \tabularnewline
\textsf{dead-ends:} & \textsf{363904522} & \textsf{872434773} & \textsf{1236339295} & \textsf{Time: 204.127 sec}\tabularnewline
\textsf{expressions:} & \textsf{25836915} & \textsf{59427019} & \textsf{85263934} & \textsf{Memory: 1351040 KB}\tabularnewline
\textsf{distinct:} & \textsf{7974910} & \textsf{9198455} & \textsf{17173365} & \tabularnewline
\textsf{Total tested:} & \textsf{(7.336e+13)} &  &  & \tabularnewline
\end{tabular}

$\vphantom{}$

The integers in braces at the right ends of result lines are the complexities,
which are computed differently from MESearch. Notice that
\begin{itemize}
\item RIES rounded the 20 digit input to 1 digit less than the 19 digit
arithmetic it is using on this computer.
\item RIES displays a candidate only if its absolute difference from the
entered float is less than that for the previously displayed candidate.
Since candidates are generated in approximate order of non-decreasing
complexity, the sequence of displayed candidates is approximately
the Pareto set of optima for the conflicting objectives of large agreement
and small complexity. Thus usually all but perhaps the last few of
the displayed candidates are possibly-useful approximations rather
than plausible candidates for the limit you seek. (But the last one
or so can be incorrect because of overfits or the limited precision
of the hardware arithmetic.)
\item The last candidate is an implicit equation with $x$ on both sides,
but it can be solved exactly for $x$ either manually or by computer
algebra to give
\begin{equation}
x=-2\pi\left(1+\dfrac{2}{-1+\cot\left(\dfrac{\pi}{2\sqrt{2}}\right)^{4}}\right)\simeq7.0895773641597344057,\label{eq:RiesExplicitSolution}
\end{equation}
which is within a few 19-digit machine-epsilons of the input float.
\item The nonequivalent candidate result implicit result \textsf{sinpi$\,$(sqrt$\,$(x)$\,$/$\,$e)
= (1/(8/phi - 1))\textasciicircum 2} is also solvable. RIES was able
to explicitly invert the backward transformation for all of the other
result lines.
\item Both of the simpler explicit solutions immediately preceding the last
output equation are equivalent to each other and to the explicit solution
(\ref{eq:RiesExplicitSolution}).\footnote{The simpler explicit equivalent results give the same 20 digit value
despite their \textsl{estimated} absolute errors being about 100 times
as large. This is probably because the estimated errors rely on linearized
error analysis and/or the limitation to hardware floating-point prevents
the float values of the nonfloat candidates from being computed more
precisely than the float input.}
\end{itemize}
RIES uses a particularly compact data structure that allows it to
build particularly large tables before having to resort to slower
secondary storage. Compiling with extended precision is slower and
reduces the maximum achievable table sizes, but the greatly increased
number of recognizable constants makes it worthwhile. To maximize
your coverage and impostor detection, compile both a 19-digit and
16-digit version if your CPU supports that, then try both.

\section{Functions built into computer algebra systems}

\subsection{The Maple $\mathrm{\mathbf{identify}}\,\mathbf{(\ldots)}$ function\label{subsec:MapleIdentify-}}

Sometimes you want to write software that invokes other software and
uses the result without human intervention for that part. For that
purpose, functions are usually better than downloadable or web applications
because function invocation from software shouldn't require human
keystrokes. Moreover, mathematical functions have a single output
and should not have side effects, making them composable, with results
that are the same regardless of temporal reorderings due to exploiting
commutativity, associativity, etc. Functions are also more suitable
for processing a batch of inputs without human intervention.

The Maple computer algebra system has a function named $\mathrm{\,identify}\,(\ldots)\,$
that can take a float as an argument and returns either one nonfloat
constant or the input float if the function cannot determine a nonfloat
that the function judges sufficiently likely to be the limit you seek.
This function is an adaptation by Kevin Hare of a Maple application
by Alan Meichsner \cite{Meichsner,KevinHare}, based on the PSLQ integer
relation algorithm.

Default arbitrary-precision Maple floats have only 10-digit significands.
Therefore you must almost always use more precision to have a reasonable
chance of success.

The fact that $\mathrm{identify}\,(\ldots)$ results are either the
input float or a nonfloat that has very nearly the same value enables
$\mathrm{identify}\,(\ldots)$ to map automatically over the parts
of non-real constants and over non-constant expressions, trying to
replace all or at least some of any floats therein with close nonfloat
constants. For example,
\[
\mathrm{identify}\,\left(\dfrac{0.27675082-1.0369278\,I}{8.3118730x^{2}+11.94535128y}\right)\:\rightarrow\dfrac{0.27675082-I\zeta(5)}{\sqrt{7}\pi x^{2}+4e\ln(3)y}
\]
where $I=\sqrt{-1}$ and $\zeta$ is the Riemann zeta function.

The one float that identify did not convert was a random float. In
contrast to applications that allow you to inspect alternatives and
reject them all, a function such as identify (\ldots ) automatically
replaces any number of floats in an expression by nonfloats without
your inspection and consent to their complexity and agreement. It
is therefore appropriate that the identify function is parsimonious
and therefore did not replace the random float with a candidate that
would almost certainly be an impostor.

To determine your confidence in a Maple identify result, try using
successively increasing precisions for the input floats until the
result appears to stabilize. Then, assess the agreements of the replaced
floats with their replacements by using the evalf function on each
of those replacements.\footnote{A replacement subexpression could still be an impostor, but if you
are satisfied with the float values of the replacements, then at least
the result is a satisfactory approximation to the original float,
which was not an exact representation either.}

The identify function has optional arguments that enable you to adjust
the integer relation models, and it is well worth doing so if the
defaults do not return a satisfactory result. For example, although
the default maximum degree for seeking algebraic numbers such as $\,$RootOf
$\,${[}\emph{polynomial}, \ldots{]}$\,$ is 6, you can change that.
Many algebraic numbers of interest have larger degree, so try larger
degrees if an algebraic number is plausible and you can compute your
wild float to high precision.

\subsection{The MPMath, SymPy and Sage $\mathrm{\mathbf{identify}}\mathbf{\,(\ldots)}$,
$\mathrm{\mathbf{findpoly}}\,\mathbf{(\ldots)}$, and $\mathrm{\mathbf{nsimplify}}\,(\ldots)$,
functions\label{subsec:The-MPMath,-SymPy}}

Sage is a software system freely downloadable from \url{http://www.sagemath.org/}
that includes NumPy, SciPy, matplotlib, SymPy, Maxima, GAP, FLINT,
R and other packages. The SymPy computer algebra system in turn includes
MPMath, which has an $\mathrm{identify}\,(\ldots)$ function analogous
to Maple's, except that identification of algebraic numbers is done
by a separate function named $\mathbf{\mathrm{findpoly}\,(\ldots)}$.
The SymPy system also has an $\mathrm{nsimplify}\,(\ldots)$ function
somewhat analogous to the NSimplify {[}\ldots{]} component of AskConstants
described in subsection \ref{subsec:AskConstants}.

\section{Functions and applications for CAS}

\subsection{An exhaustive search toolkit for \textsl{Mathematica}\label{subsec:RecognizeConstant}}

Andrzej Odrzywo\l ek analyzes the probability of correctly identifying
a float constant using forward exhaustive search \cite{OdrzywolekArticle}.
In support of that research he implemented a toolkit in \textsl{Mathematica}
freely downloable from
\noindent \begin{center}
\url{https://github.com/VA00/SymbolicRegressionPackage}
\par\end{center}

\noindent It includes a unidirectional RecognizeConstant function.
Your initial components can include any \textsl{Mathematica} symbolic
constant that can be evaluated to a float or any \textsl{Mathematica}
function that returns a float result for float inputs. Being unidirectional,
it is not as fast or memory efficient as MESearch and RIES.

My favorite function in the toolkit is a RandomExpression function
that returns a random expression of a requested complexity composed
of specified constants, variables, functions and operators. Being
exhaustive, it is particularly appropriate for generating examples
for which MESearch and RIES described in Subsection \ref{subsec:RIES}
are particularly successful.

Another useful function is EnumerateExpressions, which generates all
expressions of a specified complexity composed of specified components.

\subsection{AskConstants and associated functions for \textsl{Mathematica\label{subsec:AskConstants}}}

AskConstants is an application that I implemented in \textsl{Mathematica},
freely downloadable from \url{AskConstants.org} . AskConstants 5.0
has about 3000 integer relation models, and also bidirectional search
with a choice of precomputed tables for which the largest backward
table has about 5 million entries and the largest forward table has
about 15 million entries. The table lookup exploits sign and base
2 significand aliasing, with automatic de-aliasing and inversion for
close matches.\footnote{The bonus for base 2 aliasing exceeds that of base 10 aliasing because
2 divides a random reduced numerator or denominator more often than
10 does, and forward table entries of the form \emph{rational}$\times$\textsl{irrational}
only need to include positive \textsl{rational} numbers having an
odd numerator and denominator: The appropriate sign and power of 2
are recovered by the automatic dealiasing. Also, without loss of generality,
backward transformations of the form \textsl{rational} $\times$ \textsl{anyForwardTableEntry}
can limit the \textsl{rational} to being positive with an odd numerator
and denominator.} Thus ignoring overflow underflow, nonreal compositions and the aliasing
bonus, the largest tables cover about $(5\times10^{6})(15\times10^{6})=7.5\times10^{13}$
expressions. The backward tables and about half of the largest forward
table use tabulated and published constants. The other half of the
largest tables was generated by exhaustive table building with the
elementary functions and the most commonly-occurring special functions.
This hybrid combines advantages of precomputed tables such as ISC
with those of exhaustive breadth-first tables such as MESearch and
RIES.

Figure 4 shows the use of version 5.0 to propose a closed form for
the definite integral
\[
\intop_{0}^{\infty}\dfrac{dx}{\sqrt{1+x^{2}}+(1+x^{2})^{7/2}}
\]
that the \textsl{Mathematica} 12.1 Integrate function cannot determine.\footnote{I thank Daniel Lichtblau for this example.}

\begin{figure}[H]
\noindent \centering{}\caption{AskConstants proposes a formula that agrees with $3.26115\ldots$
to 24 digits }
\includegraphics[viewport=0bp 0bp 1827.5bp 955.89bp,scale=0.48]{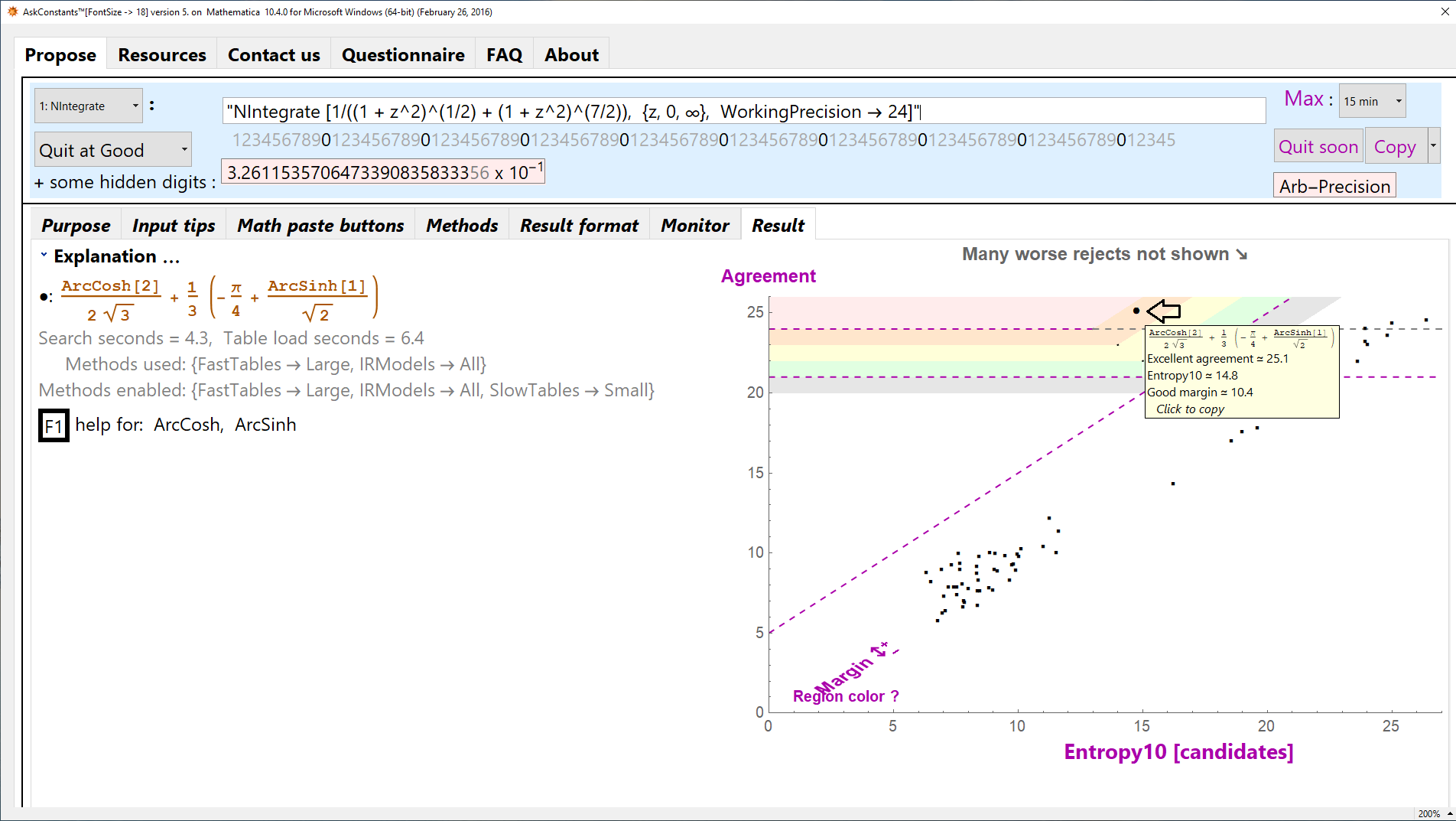}
\end{figure}

The input field near the top contains the \textsl{Mathematica} NIntegrate
function to compute the approximate float value displayed below the
digit ruler. More commonly, the numerical integration would be done
in a notebook, with the float result copied then pasted into the AskConstants
input field\label{AskConstantsScreenShot}.

AskConstants displays the candidate nonfloat
\begin{equation}
\dfrac{\mathrm{ArcCosh}\,[2]}{2\sqrt{3}}+\dfrac{1}{3}\left(-\dfrac{\pi}{2}+\dfrac{\mathrm{ArcSinh}\,[1]}{\sqrt{2}}\right).\label{eq:AcceptedCandidate}
\end{equation}
on the left side of the output and displays an associated scatter
plot of Agreement versus Entropy10 on the right side, where:
\begin{enumerate}
\item The float value of the input numerical integration is echoed under
the digit ruler, with the darker digits estimated to be correctly
rounded by significance arithmetic and with the normally undisplayed
two agreeing guard digits displayed gray.
\item The upper horizontal dashed plot line is the Precision of the input
float as estimated by significance arithmetic.
\item The \textbf{Agreement} of a candidate nonfloat is \textbf{the number
of digits that agree} with the input float within 1/2 unit, up to
the length of the input significand, including any stored digits that
are not ordinarily displayed.
\item The lower horizontal dashed line is the smallest allowable Agreement
for displaying a nonfloat on the left and plotting it as a large dot
on the scatter plot.
\item \textbf{Entropy10 }based on information theory is a measure of nonfloat
expression complexity that is the sum of the base-ten logarithms of
the absolute values of all the nonzero integers occurring in numerators
and denominators in an expression, plus about 1.0 per occurrence of
a named constant, function or operator \textendash{} counting duplicates.
The implicit units of Entropy10 are also digits.
\item It is desirable to have high agreement, but it is also desirable to
have low expression complexity because floats can be approximated
arbitrarily well by sufficiently complicated nonfloats that are impostors
rather than the limit you seek.
\item Therefore
\begin{align}
\mathrm{Margin} & :=\mathrm{Agreement-Entropy10}\label{eq:MeritDefinition-1}
\end{align}
combines these two competing objectives into one of maximizing Margin,
with the largest Margin in the upper left corner of the plot and the
smallest Margin in the lower right corner.
\item The dashed diagonal line is the smallest allowable Margin for displaying
a candidate nonfloat on the left side of the output.
\item Plot points in the upper leftmost pink region are very rarely impostors,
Plot points in the lower right white region are almost always impostors,
with other background colors having likelihoods that vary monotonically
between these extremes.
\item As illustrated, hovering the mouse over a plot point invokes a tooltip
listing the nonfloat candidate together with its Entropy10, Agreement
and Margin. The latter two include a qualitative likelihood estimate
ranging adjectives ``Excellent'' through ``Terrible'' based on
my experience with test examples.\footnote{Candidates are not produced in approximate order of increasing complexity,
so the helpful numeric likelihood estimates of MESearch are inapplicable.}
\item Hovering over a small plot point of a rejected candidate sometimes
reveals a plausible candidate for which the precision of the input
was insufficient to meet the minimum acceptable agreement and margin.
If it is the limit you seek, then increasing input precision enough
should move it vertically up into the accepted regions.
\item More generally there might be several accepted candidates. Ones judged
equivalent by applying the \textsl{Mathematica} PossibleZeroQ function
to their differences are joined by line segments, which can help you
assess them more efficiently.
\end{enumerate}
Automatic use of the backward transformation requires an inverse function
for the top-level function in the transformation; and as with most
mathematics software, \textsl{Mathematica} has very few inverse special
functions \textendash{} probably because they are extremely difficult
to implement for nonreal arguments. However, like all of the other
tools discussed in this article, AskConstants directly addresses only
real floats, and it is not too difficult to implement most inverse
special functions for real arguments and results. Consequently I did
that for about 40 special functions. Many of those are multi-branched,
such as for BesselJ, Gamma, and Zeta. Consequently I also implemented
corresponding functions that return the abscissa and ordinates of
the local infima and suprema of those functions to accomplish piecewise
monotonic partitions of the functions being inverted. The real inverse
and infima or suprema functions are separately usable, as are some
additional functions analogous to the \textsl{Mathematica} BesselJZero
function.\footnote{Some users might find these functions more frequently useful than
AskConstants.}

To overcome the limitation to constant expressions, real arguments
and results, AskConstants also contains a ProposeBestOrInput function
that maps over general expressions and over the real and imaginary
parts of non-real constants similar to the Maple identify function.

\subsubsection{NSimplify {[}...{]}}

Anyone who tests the programs described here with constants having
known nonfloat representations might eventually notice that surprisingly
often an equivalent proposed candidate is simpler than the published
expression. This is because:
\begin{itemize}
\item Manual derivations and default computer algebra simplification often
produce incompletely simplified expressions.
\item Sometimes there is no composition of builtin optional computer algebra
transformations that can produce a particularly simple representable
equivalent that exists.
\end{itemize}
Consequently the AskConstants download includes an NSimplify function
that supplements the \textsl{Mathematica} FullSimplify function by
approximating a nonfloat input expression as a float, then applying
the ProposeBestOrInput function to that float. If it returns a nonfloat
having smaller Entropy10 than the given nonfloat and the \textsl{Mathematica}
PossibleZeroQ function judges that the difference between the original
expression and proposed replacement is 0, then the original is replaced.
If that doesn't succeed, then this process is recursively applied
to subexpressions. NSimplify also tries FullSimplify so that the result
is at least that successful. This brute-force combination is slow,
but it can achieve some dramatic simplifications. Some users might
find NSimplify more frequently useful than AskConstants and ProposeBestOrInput.

As an example, applying NSimplify to
\[
\dfrac{\sqrt{\dfrac{1}{2}-\dfrac{1}{4\sqrt{\dfrac{2}{4+\sqrt{7-\sqrt{5}+\sqrt{30-6\sqrt{5}}}}}}}}{\mathrm{Root}\,\left[-97+448\,\#1-672\,\#1^{2}+560\,\#1^{3}-280\,\#1^{4}+84\,\#1^{5}-14\,\#1^{6}+\#1^{7}\,\&,1\right]}
\]
produced the result
\[
\dfrac{\sin\left(\dfrac{7\pi}{120}\right)}{2-31^{1/7}};
\]
and the optional trace reported the steps

\textbullet{} Level 2: $\mathrm{ProposeBestOrInput}\,\left[N\left[\dfrac{1}{2}-\sqrt{\dfrac{1}{4\sqrt{\ldots}}},\,26\right]\right]\mapsto\sin\left(\dfrac{7\pi}{120}\right)$,

reducing Entropy10 by 14.1; and $\,$PossibleZeroQ {[}difference{]}
$\mapsto$$\,$True.

\textbullet{} Level 2: $\mathrm{ProposeBestOrInput}\,\left[N\left[\mathrm{Root}\,\left[-97+\cdots+\#1^{7}\,\&,1\right]\right],\,38\right]\mapsto\dfrac{1}{2-31^{1/7}}$,

reducing Entropy10 by 18.6; and $\,$PossibleZeroQ {[}difference{]}
$\mapsto$$\,$True.

$\vphantom{}$

\noindent \emph{Remarks}:
\begin{enumerate}
\item NSimplify maps over over general expressions, replacing exact constant
subexpressions with less complex ones where it can.
\item The \textsl{Mathematica} PossibleZeroQ Function can incorrectly return
False, but this would merely cause NSimplify to miss an Entropy10
reduction opportunity.
\item The PossibleZeroQ function can incorrectly return True, causing NSimplify
to return a nonequivalent result. However, in this application that
also requires the parsimonious AskConstants ProposeBestOrInput function
to have generated an impostor nonfloat for a float subexpression,
and the product of these two small probabilities is such that I have
not yet noticed it happen. I suspect that the overall probability
is not much more than that of a nonequivalent result due to a bug
in the many built-in \textsl{Mathematica} functions.
\end{enumerate}

\subsection{Plouffe's inverter for Maple\label{subsec:Inverter}}

After developing ISC, Simon Plouffe developed Plouffe's Inverter \cite{PlouffesInverterArticle},
a similar Maple application with larger tables. He has several versions
using different significand sizes, and over the years his tables have
grown. As of January 2022 his largest forward table has about $8\times10^{10}$
entries. 

The same site \url{https://archive.org/} (not arXiv.org) that hosts
ISC has agreed to also store at least one of his versions for downloading.
For downloading, all three versions require you to have Maple, a fast
reliable download connection and a computer with a large amount of
secondary storage.

Therefore, I am hoping that the site will also host a free web application
alongside the ISC that they already host.

\section{Custom Integer Relation Models\label{sec:Custom-Integer-Relation}}

Wild floats often occur at the high end of a family of problems depending
on a parameter $n$, with known nonfloat values for small $n$. For
example, a noticeable number of definite integrals have nonfloat representations
that can be expressed as a rational linear combination of terms with
cofactors that are small positive integer powers of $\pi$, $\ln2$,
$\ln3$, $\zeta(m)$ with small integer $m>1$, and low-order polylogarithms
having simple arguments such as 1/2 or 1/4, perhaps multiplied by
$\sqrt{2}$ or $\sqrt{3}$. The known closed forms for small $n$
permits guesses of possible cofactors for the next value of $n$,
and if a guess is successful, then you can use that form to make cofactor
guesses for the next value of $n$, and so on.

An increasing number of computer algebra systems have builtin integer-relation
solvers. David Bailey and David Broadhurst \cite{Bailey,BaileyAndBroadhurst}
describe some particularly efficient implementations in Fortran 90
and C++. Such large models are too time consuming to make them part
of the set of more general-purpose integer relation models in tools
such as WolframAlpha, AskConstants, or the Maple and MPMath identify
functions. However, it is easy to simply invoke a built-in or stand-alone
PSLQ function with an input vector containing your float and a set
of cofactors. The LLL algorithm can also be used for this purpose
if that is available instead.

\section{The curse of extreme magnitude}

The software described here is most successful at proposing candidate
nonfloats for floats whose magnitudes are not extremely different
from 1.0. A reason for this is that representations of nonfloat constants
having extreme magnitudes often require high-complexity extreme magnitudes
of the numerators or denominators of some rational numbers therein,
which requires extreme precision input floats for integer relation
algorithms or prohibitively extreme magnitude integer subexpressions
in lookup tables. Exceptions are functions whose nonzero magnitude
can be extremely large or small for arguments that are not extremely
large or small, such as $\Gamma(98/3)\simeq8.26\times10^{34}$ or
$\mathrm{erfc}\,(9)\simeq4.14\times10^{-37}$.

Fortunately, most published \textsl{mathematics} constants do not
have extreme magnitudes. For example, in Steven Finch's table of about
10,000 such constants, nonzero magnitudes vary from 0.000111582 through
137.0359 with median about 0.9 and quartiles about $0.4$ and $1.9\,.$

\section{Some causes of impostors}
\begin{flushright}
``1.0000001 \textsl{is a crowd}''\\
\textendash{} adapted from James Thurber
\par\end{flushright}

Reasons for impostors include:
\begin{enumerate}
\item Many functions $f(x)$ have a stationary value of 1 for some value
of $x$. For example, $\cos x$, $\sec x$ and $\cosh x$, at $x=0$;
or $\tanh x$ and $\mathrm{erf}\,(x)$ as $x\rightarrow\infty$. In
a neighborhood of the stationary point, such constants that are modeled
are likely to occur as impostors for any that are not modeled, because
very low complexity $x$ can produce values very close to 1.0 and
hence each other. For example, $\sec(1/999)$ and $\cosh(1/999)$
differ by only 2 units in the 14th place. Therefore, if only one of
these was modeled, you could easily get one of them as an impostor
for the other.
\item If a sum has some terms with very small magnitudes compared to other
terms, impostors often omit those small-magnitude terms \textendash{}
at least until sufficiently large precision is used.
\item If for some relatively low complexity nonfloat constant $x$ a modeled
expression $f(x)$ agrees relatively closely with an unmodeled nonequivalent
expression $g(x)$, then that makes it easy for the modeled expression
to be an impostor for the unmodeled expression. For example the low
complexity expressions, $e^{18}$, $2\sinh18$, and $2\cosh18$ differ
by only 1 unit in the last of 16 places. If a proper subset of these
is modeled and the true limit is an unmodeled one of these, then one
of the modeled ones will almost certainly occur as an impostor having
nearly the same agreement and complexity as the correct limit.
\end{enumerate}

\section{Conclusions}

There are many good tools that can propose nonfloat candidates that
your float closely approximates. Usefully often one of those candidates
is the limit that your float would approach as the working precision
increases. Some of the tools are easy to use directly on the internet,
some are built-into a computer algebra system, and others are easy
to download and install.

Most of the programs discussed in this article can propose correct
candidates that none of the others can propose. General-purpose web
search engines, radix search, integer relation models, precomputed
tables and exhaustive run-time tables are each best at overlapping
kinds of expressions. Therefore it is worthwhile to try as many of
these programs as is reasonably convenient.

Consequently it is wise to try at least:
\begin{itemize}
\item the Online Encyclopedia of Integer Sequences\footnote{\begin{itemize}
\item This is also a good place to find high precision values and program
fragments for efficiently computing more digits.
\end{itemize}
}, 
\item at least one of the general purpose web search engines on your computers
or smart phones,
\item one of the tools that have numerous integer-relation models and can
identify algebraic numbers, preferably arbitrary degree, returning
results such as Root {[}\textsl{polynomial}, \textsl{n}{]} : AskConstants,
Maple identify, SymPy identify and FindPoly, or WolframAlpha,
\item one that has precomputed lookup tables (AskConstants, Inverse Symbolic
Calculator, and Plouffe's Inverter),
\item one that uses exhaustive bidirectional breadth-first search (MESearch
and RIES).
\end{itemize}
At the very least you should try all that are built into or downloadable
for the computer algebra systems that you already have, together with
all of the web-based tools (a browser's search engine, the Online
Encyclopedia of Integer Sequences, Inverse Symbolic Calculator and
WolframAlpha) because that is so easy to do.

You might develop favorites that tend to work better for your needs,
but it is nice to have all of these alternatives when your favorites
don't succeed and the result is important to you.

The success of your efforts depends strongly on knowing how to groom
your float for input, interpret the results, and possibly de-alias
and transform the result. These details vary greatly among the tools,
as described in this article.

When none of these programs accessible to you can propose a plausible
constant, then at least you can have the peace of mind of greatly
reducing the chance that your problem has a simple nonfloat result
that you did not find.

\section*{Acknowledgments}

Thank you Bill Gosper, Daniel Lichtblau, Robert Munafo, Simon Plouffe,
Neil Sloane and Michael Trott for your helpful suggestions.

\end{document}